\documentclass[a4paper,12pt]{article}
\def\marginpar#1{}
\input pstricks
\usepackage[latin1]{inputenc}
\usepackage[T1]{fontenc}
\usepackage{latexsym,amssymb,amsmath,epsfig}
\usepackage{a4wide}
\usepackage{graphicx}

\def\RR{\mathbb{R}}
\def\NN{\mathbb{N}}
\def\ZZ{\mathbb{Z}}
\def\II{\mathbb{I}}
\def\EE{\mathbb{E}}
\def\PP{\mathbb{P}}
\def\n3{\vert\hspace{-0.6mm}\vert\hspace{-0.6mm}\vert}

\newcommand\Cov{{\mathrm {Cov}}}
\newcommand\cov{{\mathrm {Cov}}}
\newcommand\Var{{\mathrm {Var} \,}}
\newcommand\Lip{{\mathrm {Lip } \, }}

\newtheorem{defi}{Definition}[section]
\newtheorem{lemm}[defi]{Lemma}
\newtheorem{prop}[defi]{Proposition}
\newtheorem{coro}[defi]{Corollary}
\newtheorem{theo}[defi]{Theorem}
\newenvironment{proof}{\vskip 2mm\noindent {\it Proof.}}
                    {\hfill $\square$ \vskip 2mm \noindent}

\title{A functional central limit theorem for
interacting particle systems on transitive graphs}
\author{P. Doukhan\footnote{LS-CREST, URA CNRS 220,
Paris.  \&   SAMOS-MATISSE-CES
 (Statistique Appliqu\'ee et MOd\'elisation Stochastique)
Centre d'Economie de la Sorbonne,
 Universit\'e Paris 1-Panth\'eon-Sorbonne.
90 Rue de Tolbiac, 75634 Paris Cedex 13, France. {\tt
doukhan\,@\,ensae.fr}},\,\, G. Lang\footnote{AgroParisTech, UMR MIA
518, INRA AgroParisTech, 75005 Paris France.  {\tt
gabriel.lang\,@\,agroparistech.fr} },\,\, S.
Louhichi,\footnote{Corresponding author: S. Louhichi, Universit\'e
de Paris-Sud, Probabilit\'es, statistique et mod\'elisation, B\^at.
425, 91405 Orsay Cedex, France. {\tt
sana.louhichi\,@\,math.u-psud.fr}}\,\, and B. Ycart\footnote{
LMC-IMAG,
   B.P. 53, 38041 Grenoble cedex 9, FRANCE. {\tt  Bernard.Ycart@imag.fr}}}

\begin{document}
\maketitle

\begin{abstract}
A finite range interacting particle system on a transitive graph
is considered. Assuming that the dynamics and the initial measure
are invariant, the normalized empirical distribution process
converges in distribution to a centered diffusion process.
As an application, a central
limit theorem for certain hitting times, interpreted as failure
times of a coherent system in reliability, is derived.
\end{abstract}
\vskip 1cm


\newpage
\section{Introduction}
\label{intro}
Interacting particle systems have attracted a lot of attention
because of their versatile modelling power (see for instance
\cite{Durrett99,Liggett97}). However, most available results deal
with their asymptotic behavior, and relatively few theorems
describe their transient regime. In particular, central limit
theorems for random fields have been available for a long time
\cite{Malyshev75,Neaderhouser78,Bolthausen82,Newman80,
Takahata83,CometsJanzura,ChenShao04}, diffusion approximations and
invariance principles have an even longer history
(\cite{EthierKurtz} and references therein), but those
functional central limit theorems that describe the
transient behavior of an interacting particle system are usually much
less general than their fixed-time counterparts.
Existing results (see
\cite{HolleyStrook78,HolleyStrook79,KipnisVaradhan,SethuramanXu})
require rather stringent hypotheses:
spin flip dynamics on $\mathbb{Z}$, reversibility,
exponential ergodicity, stationarity\ldots (see Holley and Strook's
discussion in the introduction of \cite{HolleyStrook79}).
The main objective of this article is to prove a functional central
limit theorem for interacting particle systems,
under very mild hypotheses, using some new
techniques of weakly dependent random fields.

Our basic reference on interacting particle systems
is the textbook by Liggett \cite{Liggett}, and
we shall try to keep our notations as close to his as possible:
$S$ denotes the (countable) set of sites, $W$ the (finite) set of
states, ${\cal X}=W^S$ the set of configurations, and
$\{\eta_t\,,\;t\geq 0\}$ an interacting particle system, i.e. a
Feller process with values in ${\cal X}$. If $R$ is a finite
subset of $S$, an empirical process is defined by counting how
many sites of $R$ are in each possible state at time $t$. This
empirical process will be denoted by $N^R = \{N^R_t\,,\;t\geq
0\}$, and defined as follows.
$$
N^R_t = (N^R_t(w))_{w\in W}\;,\quad N^R_t(w) = \sum_{x\in R}
\II_w(\eta_t(x))\;,
$$
where $\II_w$ denotes the indicator function of state $w$. Thus
$N^R_t$ is a $\NN^W$-valued stochastic process, which is not
Markovian in general. Our goal is to show that, under suitable
hypotheses, a properly scaled version of $N^R$ converges to a
Gaussian process as $R$ increases to $S$. The hypotheses will be
precised in sections \ref{basic} and \ref{covar} and the main result
(Theorem \ref{cltips2}) will be stated and proved in section
\ref{clt}. Here is a loose description of our assumptions. Dealing
with a sum of random variables, two hypotheses can be made for a
central limit theorem: weak dependence and identical distributions.
\begin{enumerate}
\item {\it Weak dependence:} In order to give it a sense, one has
to
  define a distance between sites, and therefore a graph
  structure. We shall first suppose that this (undirected) graph
  structure has bounded degree. We shall assume also finite range
  interactions: the configuration can
  simultaneously change only on a bounded set of sites, and its value
  at one site can influence transition rates only up to a fixed
  distance (Definition \ref{defetenduefinie}).
  Then if $f$ and $g$ are two functions whose dependence on
  the coordinates decreases exponentially fast with the distance from
  two distant finite sets $R_1$ and $R_2$,
  we shall prove that the covariance between
  $f(\eta_s)$ and $g(\zeta_t)$ decays exponentially fast in the
  distance between $R_1$ and $R_2$ (Proposition
  \ref{propinegcovfini}). The central limit theorem
  \ref{cltips2} will actually be proved in a much narrower setting,
  that of group invariant dynamics on a transitive graph (Definition
  \ref{definvariant}). However we believe that a covariance inequality
  for general finite range interacting particle systems is of
  independent interest. Of course the bound of Proposition
  \ref{propinegcovfini} is not uniform in time, without further
  assumptions.
\item {\it Identical distributions:} In order to ensure that the
indicator processes $\{\II_w(\eta_t(x))\,,\;t\geq 0\}$ are
identically distributed, we shall assume that the set of sites $S$
is endowed with a transitive graph structure (see
\cite{GodsilRoyle} as a general reference), and that both the
transition rates and the initial distribution are invariant by the
automorphism group action. This generalizes the notion of
translation invariance, usually considered in $\ZZ^d$
(\cite{Liggett} p.~36), and can be applied to non-lattice graphs
such as trees. Several recent articles have shown the interest of
studying random processes on graph structures more general than
$\ZZ^d$ lattices: see e.g.
\cite{Haggstrometal02,HaggstromPeres99,Haggstrometal00}, and
 for general references \cite{Peres99,Woess00}.
\end{enumerate}
Among the potential applications of our result, we chose to focus on
the hitting time of a prescribed level by a linear combination of
the empirical process. In \cite{ParoissinYcart04}, such hitting
times were considered in the application context of reliability.
Indeed the sites in $R$ can be viewed as components of a coherent
system and their states as degradation levels. Then a linear
combination of the empirical process is interpreted as the global
degradation of the system, and by Theorem \ref{cltips2}, it is
asymptotically distributed as a diffusion process if the number of
components is large. An upper bound for the degradation level can be
prescribed: the system is working as soon as the degradation is
lower, and fails at the hitting time. More precisely, let $f~:
w\mapsto f(w)$ be a mapping from $W$ to $\RR$. The total degradation
is the real-valued process $D^R=\{D^R_t\,,\;t\geq 0\}$, defined by:
$$
D^R_t = \sum_{w\in W} f(w)N^R_t(w).
$$
If $a$ is the prescribed level, the failure time of the system
will be defined as the random variable
$$
T^R_a = \inf\{t\,\geq 0\,,\; D^R_t\geq a\,\}.
$$
Under suitable hypotheses, we shall  prove that $T^R_a$ converges
weakly to a normal distribution, thus extending Theorem 1.1 of
\cite{ParoissinYcart04} to systems with dependent components.
In reliability (see \cite{BarlowProschan} for a general reference),
components of a coherent system are usually considered as
independent. The reason seems to be mathematical convenience rather
than realistic modelling. Models with dependent components have been
proposed in the setting of stochastic Petri nets
\cite{LiuChiou01,Volovoi04}. Observing that a Markovian Petri net can
also be interpreted as an interacting particle system, we believe that
the model studied here is versatile enough to be used in practical
applications.

The paper is organized as follows. Some basic facts about
interacting particle systems are first recalled in section
\ref{basic}. They are essentially those of sections I.3 and I.4 of
\cite{Liggett}, summarized here for sake of completeness, and in
order to fix notations. The covariance inequality for finite range
interactions and local functions will be given in section
\ref{covar}. Our main result, Theorem \ref{cltips2}, will be stated
in section \ref{clt}. Some examples of transitive graphs are
proposed in section \ref{graphs}. The application to hitting times
and their reliability interpretation is the object of section
\ref{hitting}. In the proof of Theorem \ref{cltips2}, we need a
spatial CLT for an interacting particle system at fixed time, i.e. a
random field. We thought interesting to state it independently in
section \ref{cltfields}: Proposition \ref{pro2r} is in the same vein
as the one proved by Bolthausen \cite{Bolthausen82} on
$\mathbb{Z}^d$, but it uses a somewhat different technique. All
proofs are postponed to section \ref{proofs}.
\section{Main notations and assumptions}
\label{basic}
In order to fix notations, we briefly recall the basic
construction of general interacting particle systems, described in
sections I.3 and I.4 of Liggett's book \cite{Liggett}.

Let $S$ be a countable set of sites, $W$ a finite set of states,
and ${\cal X}=W^S$ the set of configurations, endowed with its
product topology, that makes it a compact set. One defines a
Feller process on ${\cal X}$ by specifying the local transition
rates: to a configuration $\eta$ and a finite set of sites $T$ is
associated a nonnegative measure $c_T(\eta,\cdot)$ on $W^T$.
Loosely speaking, we want the configuration to change on $T$ after
an exponential time with parameter
$$
c_{T,\eta} = \sum_{\zeta\in W^T} c_T(\eta,\zeta).
$$
After that time, the configuration becomes equal to $\zeta$ on
$T$, with probability $c_T(\eta,\zeta)/c_{T,\eta} $. Let
$\eta^\zeta$ denote the new configuration, which is equal to
$\zeta$ on $T$, and to $\eta$ outside $T$. The infinitesimal
generator should be:
\begin{equation}
\label{defgenerateur} \Omega f(\eta) = \sum_{T\subset S}
\sum_{\zeta\in W^T} c_T(\eta,\zeta)(f(\eta^\zeta)-f(\eta)).
\end{equation}
For $\Omega$ to generate a Feller semigroup acting on continuous
functions from $X$ into $\RR$, some hypotheses have to be imposed
on the transition rates $c_T(\eta,\cdot)$.

The first condition is that the mapping $\eta\mapsto
c_T(\eta,\cdot)$ should be continuous (and thus bounded, since
${\cal X}$ is compact). Let us denote by $c_T$ its supremum norm.
$$
c_T = \sup_{\eta\in X} \,  c_{T,\eta}.
$$
It is the maximal rate of change of a configuration on $T$. One
essential hypothesis is that the maximal rate of change of a
configuration at one given site is bounded.
\begin{equation}
\label{hyp0} B = \sup_{x\in \,S}\ \sum_{T\ni\, x} c_T <\infty.
\end{equation}
If $f$ is a continuous function on ${\cal X}$, one defines
$\Delta_f(x)$ as the degree of dependence of $f$ on $x$:
$$
\Delta_f(x) = \sup\{\,|f(\eta)-f(\zeta)|\,,\;\eta,\zeta\in X
\mbox{ and } \eta(y) = \zeta(y)\; \forall\, y\neq x\,\}.
$$
Since $f$ is continuous, $\Delta_f(x)$ tends to $0$ as $x$ tends to
infinity, and $f$ is said to be {\it smooth} if $\Delta_f$ is
summable:
$$
\n3  f\n3  = \sum_{x\,\in\, S} \Delta_f(x) <\infty .
$$
It can be proved that if $f$ is smooth, then $\Omega f$ defined by
(\ref{defgenerateur}) is indeed a continuous function on ${\cal
X}$ and moreover:
$$
\|\Omega f\| \leq B \n3  f\n3  .
$$
We also need to control the dependence of the transition rates on
the configuration at other sites. If $y\in S$ is a site, and
$T\subset S$ is a finite set of sites, one defines

$$
c_T(y) = \sup \{ \, \| c_T(\eta_1,\,\cdot\,) -
c_T(\eta_2,\,\cdot\,)\|_{tv}\,,\; \eta_1(z)=\eta_2(z)\;\forall\,
z\neq y\,\} ,
$$
where $\|\,\cdot\,\|_{tv}$ is the total variation norm:
$$
\| c_T(\eta_1,\,\cdot\,) - c_T(\eta_2,\,\cdot\,)\|_{tv} =
\frac{1}{2} \sum_{\zeta \in W^T} | c_T(\eta_1,\zeta) -
c_T(\eta_2,\zeta)|.
$$
If $x$ and $y$ are two sites such that $x\neq y$, the {\it influence} of
$y$ on $x$ is defined as:
$$
\gamma(x,y) = \sum_{T\,\ni\, x} c_T(y).
$$
We will set  $\gamma(x,x)=0$ for all $x$. The influences
$\gamma(x,y)$ are assumed to be summable:
\begin{equation}
\label{hypl1} M = \sup_{x\in\, S}\ \sum_{y\in\, S} \gamma(x,y) <
\infty.
\end{equation}
Under both hypotheses (\ref{hyp0}) and (\ref{hypl1}), it can be
proved that the closure of $\Omega$ generates a Feller semigroup
$\{ S_t\,,\,t\geq 0\}$ (Theorem 3.9 p.~27 of \cite{Liggett}). A
generic process with semigroup $\{S_t\,,\,t\geq 0\}$ will be
denoted by $\{ \eta_t\,,\,t\geq 0\}$. Expectations relative to its
distribution, starting from $\eta_0=\eta$ will be denoted by
$\EE_\eta$. For each continuous function $f$, one has:
$$
S_t f(\eta) = \EE_\eta[ f(\eta_t) ] =
\EE[f(\eta_t)\,|\,\eta_0=\eta].
$$
Assume now that $W$ is ordered, (say $W=\{1,\ldots,n\}$). Let
${\cal M}$ denote the class of all continuous functions on $X$
which are monotone in the sense that $f(\eta)\leq f(\xi)$ whenever
$\eta\leq \xi$. As it was noticed by Liggett (1985) it is
essential to take advantage of monotonicity in order to prove
limit theorems for particle systems.  The following theorems
discuss a number of ideas related to monotonicity.
\begin{theo}[Theorem 2.2 Liggett, (1985)]\label{thm1}
Suppose $\eta_t$ is a Feller process on $X$ with semigroup $S(t)$.
The following statement are equivalent :
\begin{enumerate}
\item[(a)] $f\in {\cal M}$ implies $S(t)f\in {\cal M}$, for all
$t\geq 0$
\item[(b)] $\mu_1\leq \mu_2$ implies $\mu_1S(t)\leq \mu_2S(t)$ for
all $t\geq 0$.
\end{enumerate}
Recall that $\mu_1\leq \mu_2$ provided that $\int fd\mu_1\leq \int
fd\mu_2$ for any $f\in {\cal M}$.
\end{theo}
\begin{defi}
A Feller process is said to be monotone (or attractive) if the
equivalent conditions of Theorem \ref{thm1} are satisfied.
\end{defi}
\begin{theo}[Theorem 2.14 Liggett, (1985)]\label{thm2}
Suppose that $S(t)$ and $\Omega$ are respectively the semigroup
and the generator of a {\bf monotone} Feller process on $X$.
Assume further that $\Omega$ is a {\bf{bounded}} operator. Then
the following two statements are equivalent:
\begin{enumerate}
\item[(a)] $\Omega fg\geq f\Omega g+ g\Omega f$, for all $f$,
$g\in {\cal M}$
\item[(b)] $\mu S(t)$ has positive correlations whenever $\mu$
does.
\end{enumerate}
Recall that $\mu$ has positive correlation if $\int fgd\mu\geq
\left(\int fd\mu\right)\left(\int gd\mu\right)$ for any $f,g\in
\,{\cal M}$.
\end{theo}
The following corollary gives conditions under which the positive
correlation property continue to hold at later times if it holds
initially.
\begin{coro}\label{corligg}[Corollary 2.21 Liggett, (1985)]
Suppose that the assumptions of Theorem \ref{thm2} are satisfied
and that the equivalent conditions of Theorem \ref{thm2} hold. Let
$\eta_t$ be the corresponding process, where the distribution of
$\eta_0$ has positive correlations. Then for $t_1<t_2<\cdots<t_n$
the joint distribution of $(\eta_{t_1},\cdots,\eta_{t_n})$, which
is a probability measure on $X^n$, has positive correlations.
\end{coro}
\section{Covariance inequality}
\label{covar}
This section is devoted to the covariance of $f(\eta_s)$
and $g(\eta_t)$
for a finite
range interacting particle system when the underlying graph
structure has bounded degree.
Proposition \ref{propinegcovfini} shows that if $f$ and $g$
are mainly located on two finite sets $R_1$ and $R_2$, then the
covariance of $f$ and $g$ decays exponentially in the distance between
$R_1$ and $R_2$.

From now on, we assume that the set of sites $S$ is endowed with
an undirected graph structure, and we denote by $d$ the natural
distance on the graph. We will assume not only that the graph is
locally finite, but also that the degree of each vertex is
uniformly bounded.
$$
\forall x\in S\;,\quad |\{y\in S\,,\; d(x,y)=1\}|\leq r\;,
$$
where $|\,\cdot\,|$ denotes the cardinality of a finite set. Thus
the size of the sphere or ball with center $x$ and radius $n$ is
uniformly bounded in $x$, and increases at most geometrically in
$n$.
$$
|\{y\in S\,,\; d(x,y)=n\}|\leq \frac{r}{r-1}(r\!-\!1)^n
\quad\mbox{and}\quad |\{y\in S\,,\; d(x,y)\leq n\}|\leq
\frac{r}{r-2}(r\!-\!1)^n.
$$
Let $R$ be a finite subset of $S$. We shall use the following
upper bounds for the number of vertices at distance $n$, or at
most $n$ from $R$.
\begin{equation}
\label{borneboule} |\{x\in S\,,\; d(x,R)=n\}|\leq |\{y\in S\,,\;
d(x,R)\leq n\}|\leq 2|R|e^{n\rho}\;,
\end{equation}
with $\rho=\log(r-1)$.

In the case of an amenable graph (e.g. a lattice on $\ZZ^d$), the
ball sizes have a subexponential growth. Therefore, for all
$\varepsilon>0$, there exists $c$ such that~:
$$
|\{x\in S\,,\; d(x,R)=n\}|\leq |\{y\in S\,,\; d(x,R)\leq n\}|\leq
ce^{n\varepsilon}.
$$
What follows is written in the general case, using
(\ref{borneboule}). It applies to the amenable case replacing
$\rho$ by $\varepsilon$, for any $\varepsilon>0$.

We are going to deal with smooth functions, depending weakly on
coordinates away from a fixed finite set $R$. Indeed, it is not
sufficient to consider functions depending only on coordinates in
$R$, because if $f$ is such a function, then for any $t>0$, $S_tf$
may depend on all coordinates.
\begin{defi}
\label{defmainlylocated} Let $f$ be a function from $S$ into
$\RR$, and $R$ be a finite subset of $S$. The function $f$ is said
to be {\rm mainly located on} $R$ if there exists two constants
$\alpha$ and $\beta>\rho$ such that $\alpha>0$, $\beta>\rho$ and
for all $x\in \RR$:
\begin{equation}
\label{mainlylocated} \Delta_f(x) \leq \alpha e^{-\beta d(x,R)}.
\end{equation}
\end{defi}
Since $\beta>\rho$, the sum  $\sum_x \Delta_f(x)$ is finite. Therefore a
function mainly located on a finite set is necessarily smooth.

The system we are considering will be supposed to have finite
range interactions in the following sense (cf. Definition 4.17,
p.~39 of \cite{Liggett}).
\begin{defi}
\label{defetenduefinie} A particle system defined by the rates
$c_T(\eta,\cdot)$ is said to have {\it finite range interactions}
if there exists $k>0$ such that if $d(x,y)>k$:
\begin{enumerate}
\item $c_T = 0$ for all $T$ containing both $x$ and $y$\;, \item
$\gamma(x,y)=0$.
\end{enumerate}
\end{defi}
The first condition imposes that two coordinates cannot
simultaneously change if their distance is larger than $k$. The
second one says that the influence of a site on the transition
rates of another site cannot be felt beyond distance $k$.

Under these conditions, we prove the following covariance
inequality.

\begin{prop}
\label{propinegcovfini} Assume (\ref{hyp0}) and (\ref{hypl1}).
Assume moreover that the process is of finite range. Let $R_1$ and
$R_2$ be two finite subsets of $S$.
 Let $\beta$
be a constant such that $\beta>\rho$. Let $f$ and $g$ be two
functions mainly located on $R_1$ and $R_2$, in the sense that
there exist positive constants $\kappa_f,\kappa_g$ such that,
$$
\Delta_f(x) \leq \kappa_f e^{-\beta d(x,R_1)} \quad\mbox{and}\quad
\Delta_g(x) \leq \kappa_g e^{-\beta d(x,R_2)}.
$$
Then for all positive reals $s,t$,
\begin{equation}
\label{inegcovariancefini} \sup_{\eta\in X}
\Big|\cov_\eta(f(\eta_s),g(\eta_t))\Big| \leq
C\kappa_f\kappa_g(|R_1|\wedge|R_2|)e^{D(t+s)}e^{-({\beta-\rho})d(R_1,R_2)}\;,
\end{equation}
where
$$
D = 2Me^{(\beta+\rho) k} \quad\mbox{and}\quad C = \frac{2
Be^{\beta k}}{D}\left(1+\frac{e^{\rho
k}}{1-e^{-\beta+\rho}}\right).
$$
\end{prop}
\paragraph{Remark.} Shashkin \cite{Shashkin} obtains a similar
inequality for random fields indexed by $\mathbb{Z}^d$.

\

We now consider a {\it transitive graph}, such that the group of
automorphism acts transitively on $S$ (see chapter 3 of
\cite{GodsilRoyle}). Namely we need that
\begin{itemize}
    \item for any $x$ and $y$ in $S$ there exists $a$ in $Aut(S)$, such that
    $a(x)=y$.
    \item for any $x$ and $y$ in $S$ and any radius $n$, there exists $a$ in $Aut(S)$, such that
    $a(B(x,n))=B(y,n)$.
\end{itemize}
Any element $a$ of the automorphism group acts on configurations,
functions and measures on ${\cal X}$ as follows:
\begin{itemize}
\item {\it configurations:} $a\cdot \eta(x) = \eta(a^{-1}(x))$, \item
{\it functions:} $a\cdot f(\eta) = f(a\cdot \eta)$, \item {\it
measures:} $\int f\, d(a\cdot \mu) = \int(a\cdot f)\,d\mu$.
\end{itemize}
A probability measure $\mu$ on ${\cal X}$ is invariant through the
group action if $a\cdot \mu = \mu$ for any automorphism $a$, and
we want this to hold for the probability distribution of $\eta_t$
at all times $t$. It will be the case if the transition rates are
also invariant through the group action. In order to avoid
confusions with invariance in the sense of the semigroup
(Definition 1.7, p.~10 of \cite{Liggett}), invariance through the
action of the automorphism group of the graph will be
systematically referred to as ``group invariance'' in the sequel.
\begin{defi}
\label{definvariant} Let $G$ be the automorphism group of the
graph. The transition rates $c_T(\eta,\cdot)$ are said to be {\rm
group invariant} if for any $a\in G$,
$$
c_{a(T)}(a\cdot \eta,a\cdot \zeta) = c_T(\eta,\zeta).
$$
\end{defi}
This definition extends in an obvious way that of translation
invariance on $\ZZ^d$-lattices (\cite{Liggett}, p.~36).
\\
\\
{\bf{Remark.}}
 Observe
that for rates which are both  finite range and group invariant,
the hypotheses (\ref{hyp0}) and (\ref{hypl1}) 
are trivially satisfied. In that case, it is easy to check that
the semi-group $\{S_t\,,\;t\geq 0\}$ commutes with the
automorphism group. Thus if $\mu$ is a group invariant measure,
then so is $\mu S_t$ for any $t$ (see \cite{Liggett}, p.~38). In
other terms, if the distribution of $\eta_0$ is group invariant,
then that of $\eta_t$ will remain group invariant at all times.

 \section{Functional CLT}
\label{clt} Our functional central limit theorem requires that all
coordinates of the interacting particle system $\{\eta_t\,,\;t\geq
0\}$ are identically distributed.

\vskip 2mm\noindent
Let $(B_n)_{n\geq 1}$ be an increasing sequence of finite subsets of $S$
such that
\begin{equation}\label{slc}
S=\bigcup_{n=1}^{\infty}B_n,\qquad \lim_{n\rightarrow
+\infty}\frac{|\partial B_n|}{|B_n|}=0\;,
\end{equation}
recall that $|\;\cdot\,|$ denotes the cardinality and $
\partial B_n=\{x\in B_n\;,\,\exists\,y\not\in B_n,\, d(x,y)=1\}
$.
\begin{theo}
\label{cltips2} Let $\mu=\delta_{\eta}$ be  a  Dirac measure where
$\eta\in {\cal X}$ fulfills $\eta(x)=\eta(y)$ for any $x,y\in S$.
Suppose that the transition rates are group invariant. Suppose
moreover that the process is of finite range, monotone and
fulfilling the requirements of Corollary \ref{corligg}. Let
$(B_n)_{n\geq 1}$ be an increasing sequence of finite subsets of
$S$ fulfilling (\ref{slc}). Then the sequence of processes
$$
\left\{\frac{N_t^{B_n}-\EE_{\mu}N_t^{B_n}}{\sqrt{|B_n|}}\,,\;t\geq
0\right\},\qquad \mbox{ for } n=1,2,\ldots$$ converges in
${D([0,T])}$ as $n$ tends to infinity, to a centered Gaussian,
vector valued process\\ $(B(t,w))_{t\ge\,0,\,w\in\, W}$ with
covariance function $\Gamma$ defined, for $w,\, w'\in\, W$, by
$$
\Gamma_\mu(s,t)(w,w')=\sum_{x\in\, S}\cov_{\mu}\left(
\II_{w}(\eta_s(x)),\II_{w'}(\eta_t(x))\right).
$$
\end{theo}
\paragraph{Remark.} One may wonder wether such results can extend under
more general initial distributions. The point is that the covariance
inequality do not extend simply by integration with respect to
deterministic configurations. We are thankful to Pr. Penrose for
stressing our attention on this important restriction. Monotonicity
allows to get ride of this restriction.

\section{Examples of graphs}
\label{graphs}
Besides the classical lattice graphs in $\ZZ^d$ and their groups of
translations, which are considered by most authors (see
\cite{DurrettLevin94,Liggett,Liggett97}), our setting applies to a
broad range of graphs. We propose some simple examples of
automorphisms on trees, which
give rise to a large variety of non classical situations.
\vskip 2mm
The simplest example
 corresponds to regular trees defined as
follows. Consider the non-commutative free group $S$ with finite
generator set $G$. Impose that each generator $g$ is its own
inverse ($g^2=1$). Now consider $S$ as a graph, such that $x$ and
$y$ are connected if and only if there exists $g \in G$ such that
$x=yg$. Note that $S$ is a regular tree of degree equal to the
cardinality $r$ of $G$. The size of spheres is exponential:
$\left|\{y\,,\, d(x,y)=n\}\right|=r^n$. Now consider the group
action of $S$ on itself: $x\cdot y=xy$: this action is transitive
on $S$ (take $a=yx$). \vskip 2mm From this basic example it is
possible to get a large class of graphs by adding relations
between generators; for example take the tree of degree $4$,
denote by $a$, $b$, $c$, and $d$ the generators, and add the
relation $ab=c$. Then, the corresponding graph is a regular tree
of degree $4$ were nodes are replaced by tetrahedrons. The spheres
do not grow at rate $4^n$: $\left|\{y\,,\,
d(x,y)=n\}\right|=4\cdot3^{n/2}$ if $n$ is even and $\left|
\{y\,,\, d(x,y)=n\}\right|=6\cdot3^{(n-1)/2}$ if $n$ is odd.

\begin{figure}[ht]
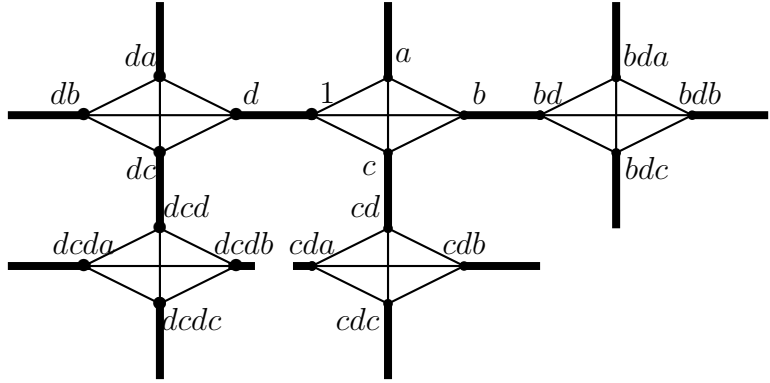

\vskip 7truecm \vbox{ \centerline{ \psset{unit=1cm}
\pspicture(2.5,2.5)(5,-0.5) \psline(1,7)(2,7.5)(3,7)(2,6.5)(1,7)
\psline(1,7)(9,7) \psline(2,6.5)(2,5.5)
\psline(1,5)(2,5.5)(3,5)(2,4.5)(1,5) \psline(2,7.5)(2,6.5)
\psline(3,7)(4,7)
\psline[showpoints=true](4,7)(5,7.5)(6,7)(5,6.5)(4,7)
\psline(6,7)(7,7)
\psline[showpoints=true](7,7)(8,7.5)(9,7)(8,6.5)(7,7)
\psline(2,4.5)(2,6) \psline(1,5)(3,5) \psline(5,7.5)(5,4.5)
\psline[showpoints=true](4,5)(5,5.5)(6,5)(5,4.5)(4,5)
\psline(4,5)(6,5) \psline(8,7.5)(8,6.5) \rput(1,7){$\bullet$}
\rput(0.75,7.3){$db$} \psline[linewidth=3pt](1,7)(0,7)
\psline[linewidth=3pt](1.09,5)(0,5)
 \rput(2,7.5){$\bullet$} \rput(1.75,7.8){$da$}
\psline[linewidth=3pt](2,7.5)(2,8.5)
 \rput(3,7){$\bullet$} \rput(3.2,7.3){$d$}
\rput(4.2,7.3){$1$} \psline[linewidth=3pt](3,7)(4,7)
\psline[linewidth=3pt](6,7)(7,7)
\psline[linewidth=3pt](2,6.5)(2,5.5)
\psline[linewidth=3pt](5,6.5)(5,5.5)
 \rput(5.2,7.8){$a$}
\psline[linewidth=3pt](5,7.5)(5,8.5) \rput(6.2,7.3){$b$}
\rput(2,6.5){$\bullet$} \rput(4.75,6.3){$c$} \rput(1.75,6.3){$dc$}
\rput(2,5.5){$\bullet$} \rput(2.35,5.8){$dcd$}
\rput(2,4.5){$\bullet$} \rput(2.41,4.3){$dcdc$}
\rput(2,4.5){$\bullet$} \rput(1,5){$\bullet$} \rput(1,5.3){$dcda$}
\rput(3,5){$\bullet$} \rput(3.1,5.3){$dcdb$}
 \psline[linewidth=3pt](2,4.5)(2,3.5)
 \rput(4.7,5.8){$cd$}\rput(4.6,4.3){$cdc$}
\psline[linewidth=3pt](5,4.5)(5,3.5)
 \rput(5.999,5.3){$cdb$}
 \psline[linewidth=3pt](6,5)(7,5)
\rput(3.999,5.3){$cda$} \psline[linewidth=3pt](4,5)(3.75,5)
\psline[linewidth=3pt](2.9,5)(3.25,5)
 \rput(4,7){$\bullet$} \rput(7.1,7.3){$bd$}
\rput(8.4,7.8){$bda$} \psline[linewidth=3pt](8,7.5)(8,8.5)
 \rput(9.1,7.3){$bdb$}
\psline[linewidth=3pt](9,7)(10,7)
 \rput(8.4,6.3){$bdc$}
 \psline[linewidth=3pt](8,6.5)(8,5.5)
\endpspicture
} } \vskip -3truecm
 \caption{Graph structure of the tree with tetrahedron
cells. The graph consists in a regular tree  of degree 4 (bold
lines), where nodes have been replaced by tetrahedrons.
Automorphisms in this graph correspond to composition of
automorphisms exchanging couples of branches of the tree (action
of generator $a$ for example) and displacements in the subjacent
regular tree.}
\end{figure}

\section{CLT for hitting times}\label{hitting}
In this section we consider the case where $W$ is ordered, the
process is monotone and satisfies the assumptions in Theorem
\ref{cltips2}, the initial condition is fixed and $f$ is an
increasing function from $W$ to $\RR$. In the reliability
interpretation, $f(w)$ measures a level of degradation for a
component in state $w$. The total degradation of the system in state
$\eta$ will be measured by the sum $\sum_{x \in B_n} f(\eta(x))$. So
we shall focus on the process $D^{(n)} = \{D_t^{(n)}\,,\,t\geq 0\}$,
where $D_t^{(n)}=D_t^{B_n}$ is the total degradation of the system
at time $t$ on the set $R=B_n$:
$$
D_t^{(n)} = \sum_{x\, \in\, B_n} f(\eta_t(x)).
$$
It is natural to consider the instants at which $D_t^{(n)}$
reaches a prescribed level of degradation. Let $k=(k(n))$ be a
sequence of real numbers. Our main object is the {\it failure
time} $T_n$, defined as:
$$
T_n = \inf\{t\geq 0\,,\; D_t^{(n)}\geq k(n)\}.
$$
In the particular case where
$W=\{\mbox{working},\,\mbox{failed}\}$ (binary components), and
$f$ is the indicator of a failed component, then $D_t^{(n)}$
simply counts the number of failed components at time $t$, and our
system is a so-called ``$k$-out-of-$n$'' system
\cite{BarlowProschan}.

Let $w_0$ be a particular state (in the reliability $w_0$ could be
the ``perfect state'' of an undergrade component). Let $\eta$ be the
constant configuration where all components are in the perfect state
$w_0$, for all $ x\in S$. Our process starts from that configuration
$\eta$, which is obviously group invariant. We shall denote by
$m(t)$ (respectively, $v(t)$) the expectation (resp., the variance)
of the degradation at time $t$ for one component.
$$
m(t) = \EE[f(\eta_t(x))\,|\,\eta_0=\eta]\;,\qquad
v(t)=\lim_{n\to\infty}\frac{\Var D_t^{(n)}}{|B_n|}.
$$
These expressions do not depend on $x\in S$, due to group
invariance.

The average degradation $D_t^{(n)}/|B_n|$ converges in probability
to its expectation $m(t)$. We shall assume that $m(t)$ is strictly
increasing on the interval $[0,\tau]$, with $0<\tau\leq+\infty$ (the
degradation starting from the perfect state increases on average).
Mathematically, one can assume that the states are ranked in
increasing order, the perfect state being the lowest. This yields a
partial order on configurations. If the rates are such that the
interacting particle system is monotone (see \cite{Liggett}), then
the average degradation increases. In the reliability
interpretation, assuming monotonicity is quite natural: it amounts
to saying that the rate at which a given component jumps to a more
degraded state is higher if its surroundings are more degraded.

We consider a ``mean degradation level'' $\alpha$, such
that $m(0)<\alpha<m(\tau)$. Assume the threshold $k(n)$ is such
that:
$$
k(n) = \alpha |B_n| + o(\sqrt{ |B_n|}).
$$
Theorem \ref{cltips2} shows that the degradation process $D^{(n)}$
should remain at distance $O(\sqrt{ |B_n|})$ from the deterministic
function $\left|B_n\right| m$. Therefore it is natural to expect
that $T_n$ is at distance $O(1/\sqrt{|B_n|})$ from the instant
$t_\alpha$ at which $m(t)$ crosses $\alpha$:
$$
t_\alpha = \inf\{t,\ m(t)=\alpha\}.
$$
\begin{theo}
\label{clthitting} Under the above hypotheses,
$$
\sqrt{ |B_n|}\left(T_n-t_\alpha\right) \xrightarrow[n\rightarrow
+\infty]{{\cal{L}}} {\cal N}(0,\sigma^2_\alpha),
$$
with:
$$
\sigma_\alpha^2 = \frac{v(t_\alpha)}{(m'(t_\alpha))^2}.
$$
\end{theo}
\section{CLT for weakly dependent random fields}
\label{cltfields}
As in section \ref{clt}, we consider a transitive graph ${\cal
G}=(S,E)$, where $S$ is the set of vertices and $E\subset
\Big\{\{x,y\},\, x,y\in\,S,\, x\neq y\Big\}$ the set of edges. For
a transitive graph, the degree $r$ of each vertex is constant (cf.
Lemma 1.3.1 in Godsil and Royle \cite{GodsilRoyle}).

For any $x$ in $S$ and for any positive integer $n$, we denote by
$B(x,n)$ the open ball of $S$ centered at $x$, with radius $n$:
$$
B(x,n)=\{y\in S,\,\,\, d(x,y)< n\}.
$$
The cardinality of the ball $B(x,n)$ is constant in $x$ and bounded as follows.
\begin{equation}\label{car}
\sup_{x\in S}|B(x,n)|\leq 2 r^n=2e^{n\rho}=: \kappa_n,
\end{equation}
where $\rho= \ln(\max(r,4)-1)$: compare with
formula (\ref{borneboule}).

Let $Y=(Y_x)_{x\in S}$ be a real valued random field. We will
measure covariances between coordinates of $Y$ on two distant sets
$R_1$ and $R_2$ through Lipschitz functions (see
\cite{DoukhanLouhichi}). A Lipschitz function is a real valued
functions $f$ defined on $\RR^n$ for some positive integer $n$,
for which
$$
\Lip f :=\sup_{x\neq y}\frac{\textstyle
\left|f(x)-f(y)\right|}{\textstyle\sum_{i=1}^n|x_i-y_i|}<\infty.
$$
We will assume the the random field $Y$ satisfies the following covariance inequality:
for any positive real $\delta$, for any
disjoint finite subsets  $R_1$ and $R_2$  of $S$ and
 for any Lipschitz functions
$f$ and $g$
  defined respectively on
$\RR^{|R_1|}$ and $\RR^{|R_2|}$, there exists a positive constant
$C_\delta$ (not depending on $f$ $g$, $R_1$ and $R_2$)
 such that
\begin{equation}\label{slcov}
\left|\Cov\left(f(Y_x,\, x\in R_1), g(Y_x,\, x\in
R_2\right)\right|\leq C_\delta \,\Lip f\,\Lip
g\,\left(|R_1|\wedge|R_2|\right)\exp\left(-\delta
d(R_1,R_2)\right).
\end{equation}
\vskip 3mm
For any finite subset $R$ of $S$, let $Z(R)=\sum_{x\in R}Y_x$. Let
$(B_n)_{n\in\NN}$ be an increasing sequence of finite subsets of
$S$ such that $|B_n|$ goes to infinity with $n$. Our purpose in
this section is to establish a central limit theorem for $Z(B_n)$,
suitably normalized. We suppose that
 $(Y_x)_{x\in S}$ is a  weakly dependent
random field according to the covariance inequality (\ref{slcov}).
\\
In Proposition \ref{pro2r} below we prove that, as in the
independent setting, a central limit theorem holds as soon as
$\Var Z(B_n)$ behaves, as $n$ goes to infinity, like $|B_n|$ (cf.
Condition (\ref{limff}) below). So the purpose of Proposition
\ref{slpro2} is to study the behavior of $\Var Z(B_n)$. We prove
that the limit (\ref{limff}) holds under two additional
conditions. The first one supposes that the cardinality of
$\partial B_n$ is asymptotically negligible compared to $|B_n|$
(cf. Condition (\ref{slc}) in section \ref{clt}); the second
condition supposes an invariance by the automorphisms of the group
${\cal G}$, of the joint distribution $(Y_x,Y_y)$ for any two
vertices $x$ and $y$. More precisely we need to have Condition
(\ref{slstationarity}) below,
\begin{equation}\label{slstationarity}
\Cov(Y_x,Y_y)=\Cov(Y_{a(x)},Y_{a(y)}),
\end{equation}
for any automorphism $a$ of ${\cal G}$.
\vskip 2mm
In order to prove Proposition \ref{pro2r}, we shall use some
estimations of Bolthausen \cite{Bolthausen82} that yield a central
limit theorem for stationary random fields on $\ZZ^d$ under mixing
conditions. Recall that the mixing coefficients used there are
defined as follows, noting by ${\cal A}_{R}$ the
$\sigma$-algebra generated by $(Y_x,\, x\in R)$,
$$
\alpha_{k,l}(n)=\sup\{|\PP(A_1\cap A_2)-\PP(A_1)\PP(A_2)|, \,
A_i\in {\cal A}_{R_i},\, |R_1|\leq k,\,
|R_2|\leq l,\, d(R_1,R_2)\geq n\},
$$
for $n\in \NN $ and $k,l\in \NN \cup {\infty}$,
$$
\rho(n)=\sup\{|\cov(Z_1,Z_2)|,\, Z_i\in L_2({\cal
A}_{\{\rho_i\}}),\, \|Z_i\|_2\leq 1,\, d(\rho_1,\rho_2)\geq n\}.
$$
Under suitable decay of $(\alpha_{k,l}(n))_n$ or of $(\rho(n))_n$,
Bolthausen \cite{Bolthausen82} proved a central limit theorem for stationary
random fields on $\ZZ^d$, using an idea of Stein. In our case,
instead of using those mixing coefficients, we describe the
dependence structure of the random fields $(Y_x)_{x\in S}$  in
terms of the gap between two Lipschitz transformations of two
disjoint blocks (the covariance inequality (\ref{slcov})
above). Those manners of describing the dependence of random
fields are quite different. As one may expect, the techniques of proof
will be different as well (see section \ref{proofs}).
\begin{prop}\label{pro2r}
 Let ${\cal G}=(S,E)$ be a transitive graph. Let
$(B_n)_{n\in\NN}$ be an increasing sequence of finite subsets of
$S$ such that $|B_n|$ goes to infinity with $n$. Let $(Y_x)_{x\in
S}$ be a real valued random field, satisfying (\ref{slcov}).
Suppose that, for any $x\in S$, $\EE Y_x=0$ and $\sup_{x\in
S}\|Y_x\|_{\infty}< \infty$.  If, there exists a finite real
number $\sigma^2$ such that
\begin{equation}\label{limff}
\lim_{n\rightarrow \infty}\frac{\Var Z(B_n)}{|B_n|}=\sigma^2,
\end{equation}
 then the quantity
$\frac{\textstyle Z(B_n)}{\textstyle \sqrt{|B_n|}}$ converges in
distribution to a centered normal law with variance $\sigma^2$.
\end{prop}
\begin{prop}\label{slpro2}
Let ${\cal G}=(S,E)$ be a transitive graph. Let $(Y_x)_{x\in S}$
be a centered real valued random field, with finite variance.
Suppose that the conditions (\ref{slcov}) and
(\ref{slstationarity}) are satisfied.
 Let $(B_n)_n$ be a sequence of finite and
increasing sets of $S$ fulfilling (\ref{slc}). Then $$\sum_{z\in
S}|\Cov(Y_0,Y_{z})|<\infty\ \ \ {\mbox{and}}\ \ \
\lim_{n\rightarrow \infty} \frac{1}{|B_n|}\Var Z(B_n)=\sum_{z\in
S}\Cov(Y_0,Y_{z}).$$
\end{prop}

\section{Proofs}
\label{proofs}
\subsection{Proof of Proposition \ref{propinegcovfini}}
Let $\Gamma$ denote the matrix $(\gamma(x,y))_{x,y\in S}$, and let
it operate on the right on the space of summable series
$\ell_1(S)$ indexed by the denumerable set $S$:
$$
u = (u(x))_{x\in S} \mapsto \Gamma u= (\Gamma u(y))_{y\in S}\,,
$$
with~:
$$
\Gamma u(y) = \sum_{x\in S} u(x)\, \gamma(x,y).
$$
(We have followed Liggett's \cite{Liggett} choice of denoting by $\Gamma
u$ the product of $u$ by $\Gamma$ on the right.)
Thanks to hypothesis (\ref{hypl1}), this defines a bounded
operator of $\ell_1(S)$, with norm $M$. Thus for all $t\geq 0$,
the exponential of $t\Gamma$, is well defined, and gives another
bounded operator of $\ell_1(S)$:
$$
\exp(t\Gamma)u = \sum_{n=0}^\infty \frac{t^n \Gamma^n u}{n!}.
$$
If $f$ is a smooth function, then $\Delta_f =(\Delta_f(x))_{x\in
S}$, is an element of $\ell_1(S)$. Applying $\exp(t\Gamma)$ to
$\Delta f$ provides a control on $S_tf$ as shows the following
proposition (cf. Theorem 3.9 of  \cite{Liggett}).
\begin{prop}
\label{propsmooth} Assume {\rm (\ref{hyp0})} and {\rm (\ref{hypl1})}. Let $f$
be a smooth function. Then,
\begin{equation}
\label{ineg1} \Delta_{S_tf} \leq \exp(t\Gamma)\Delta_f .
\end{equation}
\end{prop}
It follows immediately that if $f$ is a smooth function then
$S_tf$ is also smooth and:
$$
\n3  S_t f\n3  \leq e^{tM}\,\n3  f\n3 \;,
$$
because the norm of $\exp(t \Gamma)$ operating on $\ell_1(S)$ is
$e^{tM}$. \\
A similar bound for covariances will be our starting point (cf.
Proposition 4.4, p.~34 of \cite{Liggett}).
\begin{prop}
\label{inegcovbasique} Assume {\rm (\ref{hyp0})} and {\rm (\ref{hypl1})}. Then
for any smooth functions $f$ and $g$ and for all $t\geq 0$, one
has,
\begin{equation}
\label{inegcovariance} \| S_tfg - (S_tf)(S_tg)\| \leq \sum_{y,z\in
S} \left(\sum_{T\ni y,z} c_T\right) \int_0^t (\exp(\tau
\Gamma)\Delta_f)(y) (\exp(\tau \Gamma)\Delta_g)(z)\,d\tau.
\end{equation}
\end{prop}
In terms of the process $\{\eta_t\,,\,t\geq 0\}$, the left member
of (\ref{inegcovariance}) is the uniform bound for the covariance
between $f(\eta_t)$ and $g(\eta_t)$.
$$
\| S_tfg - (S_tf)(S_tg)\| = \sup_{\eta\in X} \Big|
\EE_\eta[f(\eta_t)g(\eta_t)] -
\EE_\eta[f(\eta_t)]\EE_\eta[g(\eta_t)] \Big|.
$$
A slight modification of (\ref{inegcovariance}) gives a bound on
the covariance of $f(\eta_s)$ with $g(\eta_t)$, for $0\leq s\leq
t$. From now on, we shall denote by $\cov_\eta$ covariances
relative to the distribution of $\{\eta_t\,,\;t\geq 0\}$, starting
at $\eta_0=\eta$:
$$
\cov_\eta(f(\eta_s),g(\eta_t)) = \EE_\eta[f(\eta_s)g(\eta_t)] -
\EE_\eta[f(\eta_s)]\EE_\eta[g(\eta_t)].
$$
\begin{coro}
\label{corocovariancest} Assume {\rm (\ref{hyp0})} and {\rm (\ref{hypl1})}.
Let $f$ and $g$ be two smooth functions. Then for all $s$ and $t$
such that $0\leq s\leq t$,
\begin{equation}
\label{inegcovariancest} \sup_{\eta\in X}
\Big|\cov_\eta(f(\eta_s),g(\eta_t))\Big| \leq \sum_{y,z\in S}
\left(\sum_{T\ni y,z} c_T\right) \int_0^s
(\exp(\tau\Gamma)\Delta_f)(y)(\exp(\tau\Gamma)\Delta_{S_{t-s}g})(z)\,d\tau.
\end{equation}
\end{coro}
{\bf{Proof of Corollary \ref{corocovariancest}.}}  We have, using
the semigroup property,
$$
\EE_\eta[ f(\eta_s) g(\eta_t)] =
\EE_\eta[f(\eta_s)\EE[g(\eta_t)\,|\,\eta_s]] = \EE_\eta[f(\eta_s)
S_{t-s}g(\eta_s)] = S_s(fS_{t-s}g)(\eta).
$$
Also,
$$
\EE_\eta[g(\eta_t)] = S_tg(\eta) = S_s(S_{t-s}g)(\eta).
$$
Applying (\ref{inegcovariance}) at time $s$ to $f$ and $S_{t-s}g$,
yields the result.\,\, $\Box$
\vskip 3mm
 In order to apply
(\ref{inegcovariancest}) to functions mainly located on finite
sets, we shall need to control the effect of $\exp(t\Gamma)$ on a
sequence $(\Delta_f(x))$ satisfying (\ref{mainlylocated}). This
will be done through the following technical lemma.
\begin{lemm}
\label{lemmetechniquegamma} Suppose that the process is of finite
range. Let $R$ be a finite
set of sites. Let $u=(u(x))_{x\in S}$ be an element of
$\ell_1(S)$. If for all $x\in S$, $u(x)\leq \alpha e^{-\beta
d(x,R)}$, with $\alpha>0$ and $\beta>\rho$, then for all $y\in S$,
$$
|(\exp(t\Gamma)u)(y)| \leq \alpha\exp(2tMe^{(\beta+\rho)
k})\,e^{-\beta d(y,R)}.
$$
\end{lemm}
This lemma, together with Proposition \ref{propsmooth}, justifies
Definition \ref{defmainlylocated}. Indeed, if $f$ is mainly
located on $R$, then by (\ref{ineg1}) and Lemma
\ref{lemmetechniquegamma}, $S_tf$ is also mainly located on $R$,
and the rate of exponential decay $\beta$ is the same for both
functions.
\\
{\bf{Proof of Lemma \ref{lemmetechniquegamma}.}} Recall that
$$
\Gamma u(y) = \sum_{x\in S} u(x)\gamma(x,y).
$$
Observe that if $\gamma(x,y)>0$, then the distance from $x$ to $y$
must be at most $k$ and thus the distance from $x$ to $R$ is at
least $d(y,R)-k$. If $u(x)\leq \alpha e^{-\beta d(x,R)}$ then:
$$
\Gamma u(y) \leq 2\alpha e^{\rho k}e^{-\beta(d(y,R)-k)}M = 2\alpha
e^{(\beta +\rho)k} M e^{-\beta d(y,R)}.
$$
Hence by induction,
$$
\Gamma^n u(y) \leq \alpha 2^ne^{(\beta+\rho) kn} M^{n} e^{-\beta
d(y,R)}.
$$
The result follows immediately.\,\, $\Box$
\vskip 3mm
Together with (\ref{inegcovariancest}), Lemma
\ref{lemmetechniquegamma} will be the key ingredient in the proof
of our covariance inequality.
\\
{\bf{End of the proof of Proposition \ref{propinegcovfini}.}}
Being mainly located on finite sets, the functions $f$ and $g$ are
smooth. By (\ref{inegcovariancest}), the covariance of $f(\eta_s)$
and $g(\eta_t)$ is bounded by $M(s,t)$ with:
$$
M(s,t)= \sum_{y,z\in\, S} \left(\sum_{T\ni\, y,z} c_T\right)
\int_0^s
(\exp(\tau\Gamma)\Delta_f)(y)(\exp(\tau\Gamma)\Delta_{S_{t-s}g})(z)\,d\tau.
$$
Let us apply Lemma \ref{lemmetechniquegamma} to $\Delta_f$ and
$\Delta_{S_{t-s}g}$.
\begin{equation}
\label{maj1} (\exp(\tau\Gamma)\Delta_f)(y) \leq \kappa_f\exp(\tau
Me^{(\beta+\rho) k})
  e^{-\beta d(y,R_1)}
= \kappa_f e^{D\tau}e^{-\beta d(y,R_1)}.
\end{equation}
The last bound, together with (\ref{ineg1}), gives
$$
\Delta_{S_{t-s}g}(x)\leq (\exp((t-s)\Gamma)\Delta_g)(x) \leq
\kappa_g e^{D(t-s)}e^{-\beta d(x,R_2)}.
$$
Therefore~:
\begin{equation}
\label{maj2} (\exp(\tau\Gamma)\Delta_{S_{t-s}g})(z)\leq \kappa_g
e^{D(\tau+t-s)} e^{-\beta d(z,R_2)}.
\end{equation}
Inserting the new bounds (\ref{maj1}) and (\ref{maj2}) into
$M(s,t)$, we obtain
$$
M(s,t)\leq \sum_{y,z\in\, S} \left(\sum_{T\ni\, y,z} c_T\right)
\kappa_f\kappa_g e^{-\beta(d(y,R_1)+d(z,R_2))} \int_0^s
e^{D(2\tau+t-s)}\,d\tau.
$$
Now if $d(y,z)>k$ and $y,z\in T$, then $c_T$ is null by Definition
\ref{defetenduefinie}. Remember moreover that by hypothesis
(\ref{hyp0}):
$$
B=\sup_{u\in S} \sum_{T\ni u} c_T  <\infty.
$$
Therefore~:
\begin{equation}\label{avantderniere}
M(s,t) \leq \kappa_f\kappa_g\frac{Be^{D(s+t)}}{2D}\sum_{y\in S}
\sum_{d(y,z)\leq k} e^{-\beta(d(y,R_1)+d(z,R_2))}.
\end{equation}
In order to evaluate the last quantity, we have to distinguish two
cases. \\
\\
$\bullet$ If $d(R_1,R_2)\leq k$, then
\begin{eqnarray*}
\sum_{y\in S} \sum_{d(y,z)\leq k}
e^{-\beta(d(y,R_1)+d(z,R_2))}&\leq& 2e^{\rho k}\sum_{y\in S}
e^{-\beta d(y,R_1)}
\\
&\leq & 2e^{\rho k}\sum_{n\in\NN}\sum_{y\in S} e^{-\beta
d(y,R_1)}\II_{d(y,R_1)=n}
\\
&\leq& 4|R_1|e^{\rho k}\sum_{n=0}^\infty e^{(\rho-\beta )n}
\\
&\leq &  \frac{4|R_1| e^{\rho k}}{1-e^{-(\beta-\rho)}}\\
& \leq & |R_1|\frac{4 e^{(\rho+\beta)
k}}{1-e^{-(\beta-\rho)}}e^{-\beta d(R_1,R_2)}\\
& \leq & |R_1|\frac{4 e^{(\rho+\beta)
k}}{1-e^{-(\beta-\rho)}}e^{-(\beta - \rho) d(R_1,R_2)}
\end{eqnarray*}
$\bullet$ If $d(R_1,R_2)>k$, then we have, noting that
$d(y,R_1)+d(z,R_2)\geq d(R_1,R_2)-d(y,z)$ and that $d(y,z)\leq k$,

\begin{eqnarray*}
{\lefteqn{\sum_{y\in S} \sum_{d(y,z)\leq k}
e^{-\beta(d(y,R_1)+d(z,R_2))}}}\\
&&\leq \sum_{d(y,R_1)\leq d(R_1,R_2)-k} \sum_{d(y,z)\leq
k}e^{-\beta (d(R_1,R_2)-k)} +\sum_{d(y,R_1)\geq d(R_1,R_2)-k}
\sum_{d(y,z)\leq k}e^{-\beta d(y,R_1)}
\\[2ex]
&&\leq 4 |R_1|\,e^{\rho (d(R_1,R_2)-k)}e^{\rho k}e^{-\beta
(d(R_1,R_2)-k)} +4|R_1|e^{\rho k}\sum_{n\geq d(R_1,R_2)-k}
e^{(\rho-\beta )n}
\\[2ex]
&&\leq 4|R_1|\, e^{\beta
k}\left(1+\frac{1}{1-e^{-(\beta-\rho)}}\right)e^{-(\beta-\rho)d(R_1,R_2)}.
\end{eqnarray*}

By inserting the latter bound into (\ref{avantderniere}), one
obtains,
$$
M(s,t)\leq
C\kappa_f\kappa_g|R_1|e^{D(t+s)}e^{-({\beta-\rho})d(R_1,R_2)}\;,
$$
with~:
$$
C = \frac{2B}{D}e^{\beta k}\left(1+\frac{e^{\rho
k}}{1-e^{-\beta+\rho}}\right).\qquad \Box
$$
The covariance inequality (\ref{inegcovariancefini}) implies that
the covariance between two functions essentially located on two
distant sets decays exponentially with the distance of those two
sets, whatever the instants at which it is evaluated. However the
upper bound increases exponentially fast with $s$ and $t$. In the
case where the process $\{\eta_t\,,\;t\geq 0\}$ converges at
exponential speed to its equilibrium, it is possible to give a
bound that increases only in $t-s$, thus being uniform in $t$ for
the covariance at a given instant $t$.

\subsection{Proof of Theorem  \ref{cltips2}}
\subsubsection{Finite dimensional laws}
Let ${\cal G}=(S,E)$ be a transitive graph and $Aut({\cal G})$ be
the automorphism group of ${\cal G}$. Let $\mu$ be a probability
measure on ${\cal X}$ invariant through the automorphism group
action. Let $(\eta_t)_{t\geq 0}$ be an interacting particle system
fulfilling the requirements of Theorem \ref{cltips2}. Recall that
$\{S_t\,,\;t\geq 0\}$ denotes the semigroup and $\mu S_t$ the
distribution of $\eta_t$, if the distribution of $\eta_0$ is $\mu$.
\begin{prop}\label{fidi0}
Let $(B_n)_n$ be an increasing sequence of finite subsets of $S$
fulfilling {\rm (\ref{slc})}. Let assumptions of   Theorem
\ref{cltips2} hold.
 Then for any fixed positive real numbers $t_1\leq
t_2\leq\cdots\leq t_k$, the random vector
$$\frac{1}{\sqrt{|B_n|}}\left(N_{t_1}^{B_n}-\EE_{\mu} N_{t_1}^{B_n},
N_{t_2}^{B_n}-\EE_{\mu} N_{t_2}^{B_n}
,\ldots,N_{t_k}^{B_n}-\EE_{\mu} N_{t_k}^{B_n}\right)$$ converges
in distribution, as $n$ tends to infinity, to a centered Gaussian
vector with covariance matrix $(\Gamma_{\mu}(t_i,t_j))_{1\leq
i,j\leq k}$.
\end{prop}
{\bf{Proof of Proposition \ref{fidi0}.}} We will only study the
convergence in distribution of the vector
$$\frac{1}{\sqrt{|B_n|}}\left(N_{t_1}^{B_n}-\EE_{\mu} N_{t_1}^{B_n},
N_{t_2}^{B_n}-\EE_{\mu} N_{t_2}^{B_n}\right)\;,$$ the general
case being similar.
 For $i=1,2$, we denote by $\alpha_i=(\alpha_i(w))_{w\in W}$ two
 fixed vectors of $\RR^{|W|}$. We have, denoting by $\cdot$ the usual
 scalar product,
 \begin{eqnarray*}
{\lefteqn{\frac{1}{\sqrt{|B_n|}}\sum_{i=1}^2 \alpha_i \cdot
\left(N_{t_i}^{B_n}-\EE_{\mu}
N_{t_i}^{B_n}\right)}}\\
&& = \frac{1}{\sqrt{|B_n|}}\sum_{x\in
B_n}\left(\sum_{i=1}^2\left(\sum_{w\in W}
\alpha_i(w)(\II_w(\eta_{t_i}(x))-\PP_{\mu}(\eta_{t_i}(x)=w))\right)\right)\\
&&= \frac{1}{\sqrt{|B_n|}}\sum_{x\in B_n}Y_x,
\end{eqnarray*}
 where $(Y_x)_{x\in S}$ is the random field defined by
\begin{equation}\label{Ydef}
Y_x = \sum_{i=1}^2\left(\sum_{w\in W}
\alpha_i(w)(\II_w(\eta_{t_i}(x))-\PP_{\mu}(\eta_{t_i}(x)=w))\right)=:F_1(\eta_{t_1}(x))
+ F_2(\eta_{t_2}(x)).
\end{equation}
The purpose is then to prove a central limit theorem for the sum
$\sum_{x\in B_n}Y_x$. For this, we shall study the nature of the
dependence of $(Y_x)_{x\in S}$.

Let $R_1$ and $R_2$ be two finite and disjoints subsets of $S$.
Let $k_1$ and $k_2$ be two
 real valued functions defined respectively on
$\RR^{|R_1|}$ and $\RR^{|R_2|}$. Let
  $K_1$, $K_2$ be two real valued functions,
defined respectively on $W^{R_1}$ and $W^{R_2}$, by
$$K_j(\nu,\eta)=k_j(F_1(\nu(x))+ F_2(\eta(x)),\ x\in R_j),\ \
\ j=1,2. $$ 
Let ${\cal L}$ be the class of real valued Lipschitz functions $f$ defined
on $\RR^n$, for some positive integer $n$, for which
$$
\Lip f :=\sup_{x\neq y}\frac{\textstyle
\left|f(x)-f(y)\right|}{\textstyle\sum_{i=1}^n|x_i-y_i|}<\infty.
$$
We assume that $k_1$ and $k_2$ belong to ${\cal L}$.
  Recall that
\begin{eqnarray*}
\Cov_{\eta}(k_1(Y_x,\, x\in R_1), k_2(Y_x,\,x \in
R_2))&=&\Cov_{\eta}\left(K_1(\eta_{t_1},\eta_{t_2}),
K_2(\eta_{t_1},\eta_{t_2})\right) \\
\end{eqnarray*}
But
$$|K_1(\eta_{t_1},\eta_{t_2})-K_1(\eta'_{t_1},\eta_{t_2})|\leq
4\Lip k_1\sum_{w\in W}|\alpha_1(w)|\sum_{x\in
R_1}|\eta_{t_1}(x)-\eta'_{t_1}(x)|
$$
Denote $ A_1(W)=4\Lip k_1\sum_{w\in W}|\alpha_1(w)|$. Then, the
functions
$$
\eta_{t_1}\longrightarrow (\Lip k_1)A_1(W) \sum_{x\in
R_1}\eta_{t_1}(x) \pm\,\,K_1(\eta_{t_1},\eta_{t_2})
$$
are increasing. Hence, the functions
$$
G_1^{\pm}:\,(\eta_{t_1},\eta_{t_2})\longrightarrow \Lip k_1
\sum_{x\in R_1}\left(A_1(W)\eta_{t_1}(x)+A_2(W)\eta_{t_2}(x)\right)
\pm\,\,K_1(\eta_{t_1},\eta_{t_2})
$$
are increasing  coordinate by coordinate. This also holds for,
$$
G_2^{\pm}:\,(\eta_{t_1},\eta_{t_2})\longrightarrow \Lip k_2
\sum_{x\in R_2}(A_1(W)\eta_{t_1}(x)+A_2(W)\eta_{t_2}(x))
\pm\,\,K_2(\eta_{t_1},\eta_{t_2}).
$$
Under assumptions of Theorem \ref{thm2} and of its Corollary
\ref{corligg}, the vector $(\eta_{t_1},\eta_{t_2})$ has positive
correlation so that
$$
\Cov_{\eta}(G_1^{\pm}(\eta_{t_1},\eta_{t_2}),
G_2^{\pm}(\eta_{t_1},\eta_{t_2}))\geq 0.
$$
This gives
\begin{eqnarray*}
&&\left|\Cov_{\eta}(k_1(Y_x,\, x\in R_1), k_2(Y_x,\,x \in
R_2))\right|\\
&& \leq \Lip k_1\Lip k_2 \sum_{x\in R_1}\sum_{y\in
R_2}\Cov_{\eta}(A_1(W)\eta_{t_1}(x)+A_2(W)\eta_{t_2}(x),A_1(W)\eta_{t_1}(y)+A_2(W)\eta_{t_2}(y)).
\end{eqnarray*}
 From this bilinear formula, we now apply Proposition
\ref{propinegcovfini} and  obtain the following covariance
inequality: for finite subsets $R_1$ and $R_2$ of $S$, we have
letting $\delta=\beta-\rho$,
$$\left|\Cov_{\eta}\left(K_1(\eta_{t_1},\eta_{t_2}),K_2(\eta_{t_1},\eta_{t_2})\right)\right|\leq  C_{\delta} \Lip k_1\Lip
k_2\left(|R_1|\wedge|R_2|\right)\exp\left(-\delta
d(R_1,R_2)\right),$$ where $C_{\delta}$ is a positive constant
depending on $\beta$ and not depending on $R_1$, $R_2$, $k_1$ and
$k_2$.
\\
We then deduce from Proposition \ref{pro2r} that
$\frac{1}{\sqrt{|B_n|}}\sum_{x\in B_n}Y_x$ converges in distribution
to a centered normal law as soon as the quantity
$\Var_{\mu}(\sum_{x\in B_n}Y_x)/|B_n|$ converges as $n$ tends to
infinity to a finite number $\sigma^2$. This variance converges if
the requirements of Proposition \ref{slpro2} are satisfied. For
this, we first check the condition of invariance
(\ref{slstationarity}):
$$
\cov_{\mu}(Y_x,Y_y)=\cov_{\mu}(Y_{a(x)},Y_{a(y)}),
$$
for any automorphism $a$ of ${\cal G}$ and for $Y_x$ as defined by
(\ref{Ydef}). We recall that the initial distribution  is a Dirac
distribution on the configuration $\eta$. Then it has positive
correlations. We have supposed that $\eta(x)=\eta(y)$ for all
$x,y\in S$, hence $a\cdot \mu=\mu$ and the group invariance
property of the transition rates proves that $\mu=\delta_{\eta}$
fulfills (\ref{essai}) below and then (\ref{slstationarity}) will
hold. Condition (\ref{essai}) is true thanks to the following
estimations valid for any suitable real valued functions $f$ and
$g$,
\begin{eqnarray}\label{essai}
{\lefteqn{\EE_{\mu}(f(\eta_{t_1})g(\eta_{t_2}))}}
{\nonumber}\\
&& =\int d\mu(\eta)S_{t_1}\left(fS_{t_2-t_1}g\right)(\eta)
{\nonumber}
\\
&&= \int d\mu(\eta)\, a\cdot
S_{t_1}\left(fS_{t_2-t_1}g\right)(\eta)\ \ \ {\mbox{since}}\ \ \ \
a\cdot\mu=\mu {\nonumber}
\\
&&=\int d\mu(\eta) S_{t_1}\left((a\cdot f)S_{t_2-t_1}(a\cdot
g)\right)(\eta)\
\ \ {\mbox{since}}\ \ \ \ a\cdot (S_sf)=S_s(a\cdot f) {\nonumber}\\
&& = \EE_{\mu}((a\cdot f)(\eta_{t_1})(a\cdot
g)(\eta_{t_2}))=\EE_{\mu}(f(a\cdot\eta_{t_1})g(a\cdot\eta_{t_2})).
\end{eqnarray}
 Hence Proposition \ref{slpro2} applies and gives
\begin{eqnarray*}
{\lefteqn{\sigma^2=\sum_{z\in S}\cov_{\mu}(Y_0,Y_z)}} \\
&& = \sum_{i,j=1}^2\sum_{w,w'\in W}\alpha_i(w)\alpha_j(w')\sum_{z\in
S}\cov_{\mu}\left(\II_w(\eta_{t_i}(0)),\II_{w'}(\eta_{t_i}(z))\right)
\\
&&= \sum_{i,j=1}^2 \alpha_i^t \Gamma_{\mu}(t_i,t_j) \alpha_j,
\end{eqnarray*}
where $\Gamma_{\mu}(t_i,t_j)$ is the covariance matrix as defined in
Theorem \ref{cltips2}; with this we complete the proof of
Proposition \ref{fidi0}.

\subsubsection{Tightness}\label{secwdep}

First we establish covariance inequalities for the counting
process. Denote $g_{s,t,w}(\eta,y)=\II_w(\eta_{
t}(y))-\II_w(\eta_{s}(y))$ and for any multi-index ${\bf
y}=(y_1,\ldots,y_u)\in S^u$, for any  state vector ${\bf
w}=(w_1,\ldots, w_u)\in W^u$, $\Pi_{{\bf y},{\bf w}}=\prod_{
\ell=1}^ug_{s,t,w_\ell}(\eta,y_\ell)$.
 Following (\ref{inegcovariancefini}), for $\beta>\rho$, for any
$r$-distant finite multi-indices ${\bf y}\in S^u$ and ${\bf z}\in
S^v$ ,
 for any times $0\le s\le t\leq T$ and
 for any  state vectors ${\bf w}\in W^u$
and $ {\bf w'}\in W^v$
\begin{equation}\label{inegcovsimp}
\left|\cov_\eta\left(\Pi_{{\bf y},{\bf w}},\Pi_{{\bf z},{\bf w'}}
\right)\right|\leq 4C (u\wedge v)e^{2DT }e^{-(\beta-\rho) r}\equiv
c_0  (u\wedge v)e^{-c r} ,
\end{equation}
for $c=\beta-\rho$ and $c_0=\frac{\textstyle
4Be^{2DT}e^{-(\beta-\rho)r}(2-e^{-c})}{\textstyle Me^{\rho
k}(1-e^{-c})}$.
\begin{lemm}  \label{cor:covx}There exist $\delta_0>0$ and $K_\Omega>0$ such that for
$|s-t|<\delta_0$:
\begin{equation}\label{eqn:covx1}
\left|\cov_\eta\left(\Pi_{{\bf x},{\bf w}},\Pi_{{\bf y},{\bf w'}}
\right)\right|\leq K_\Omega |t-s|.
\end{equation}
\end{lemm}

\begin{proof} Denote $f(\eta)=\II_w(\eta(x))$ then
$g_{t+h,t,w}(\eta,x)=S_hf(\eta_{t})-f(\eta_t)$; the properties of
the generator $\Omega$ imply that
$$
\lim_{h\rightarrow 0} \frac{S_{h}f(\eta)-f(\eta)}h= \Omega f(\eta)
$$
But
\begin{eqnarray*}
|\Omega f(\eta)|&\leq& \sum_{T\subset S}\sum_{\zeta \in
W^T}c_T(\eta,\zeta)|f(\eta^\zeta)-f(\eta)|
\\ &\leq&\sum_{T\subset S, x \in T}c_T(\eta)\leq \sum_{T\subset S, x \in T}c_T\leq  C_\Omega
\end{eqnarray*}
so that for $h>0$ tending to zero
$$
|g_{s,s+h,w}(\eta,x)|\leq C_\Omega h+o(h)
$$
Because $\Omega$ is group invariant, the remainder term is uniform
with respect to index $x$, so that we find convenient $\delta_0$
and $K_\Omega$ uniformly with respect to location.
\end{proof}
\noindent From inequality (\ref{inegcovsimp}) and lemma
\ref{cor:covx}, we deduce the following moment inequality:
\begin{prop}\label{pro:moment}
Choose $l$ and $c$ such that $\rho(2l-1)<c$. For $(s,t)$ such that
$|t-s|<\delta_0\wedge c_0e^c/K_\Omega$:
\begin{multline}
\EE  (N_t^{B_n}-N_s^{B_n})^{2l}\leq\frac{(4l-2)!(c_0e^
{2c})^{\frac{\rho l}c}}{(2l)!(2l-1)!}
\\\left(\frac{2^{2l}(2l)!(c_0e^{ 2c})^{\frac{\rho (l-1)}c}}{c_1}
|B_n|^{1-l}(K_\Omega|t-s|)^{1-\frac{\rho (2l-1)}c}
 +\left(\frac8 {c_1}\right)^{l}(K_\Omega|t-s|)^{l-\frac {\rho l} c}
\right),
\end{multline}
where $c_1=\rho \wedge (c-\rho(2l-1))$.
\end{prop}

\begin{proof} Recall  that $
N_t^{B_n}-N_s^{B_n}=\frac{1}{\sqrt{|B_n|}}\sum_{x\in B_n }
g_{s,t,w}(\eta,x)$.  Note that  the value of $\Pi_{\bf x}$  does
not depend on the order of the elements $x_1,\ldots,x_L.$ The
index $\bf x$ is said to split into ${\bf y}=(y_1,\ldots, y_M)$
and ${\bf z}=(z_1,\ldots,z_{L-M})$ if one can write
$y_1=x_{\sigma(1)},\ldots,y_M=x_{\sigma(M)}$ and
$z_1=x_{\sigma(M+1)},\ldots,z_{L-M}=x_{\sigma(L)}$ for some
bijection $\sigma:\{1,\ldots, L\}\to\{1,\ldots, L\}$. We adapt
lemma 14 in Doukhan \& Louhichi  \cite{DoukhanLouhichi} to the
series $(g_{t,s,w}(\eta,x))_{x\in B_n}$. For any integer $q\geq
1$, set :
\begin{equation}\label{eqn:defaq}
A_q(n)=\sum_{{\bf x}\in B_n^q} \left|\EE \Pi_{\bf x,w}\right|,
\end{equation}
then,
\begin{equation}\label{eqn:201}
\EE (N_s^{B_n}-N_t^{B_n})^{2l}\leq |B_n|^{-l}A_{2l}(n).
\end{equation}
If $q\geq 2$, for a multi-index ${\bf x}=(x_1,\ldots,x_q)$ of
elements of $S$, the gap is defined by the maximum of the integers
$r$ such that the index may split into two non-empty sub-indices
${\bf y}=( y_1,\ldots,y_h)$ and ${\bf z}=(z_1,\ldots,z_{q-h})$
whose mutual distance equals $r$: $d({\bf y(x)},{\bf
z(x)})=\min\{d(y_a,z_b);1\le a\le h,1\le b\le q-h \}=r$. If the
sequence is constant, its gap is $0$. Define the set $G_r(q,n) =\{
{\bf x}\in B_n^q$ and the gap of ${\bf x}$ is $r$\}. Sorting the
sequences of indices by their gap:
\begin{eqnarray}
A_q(n)&\leq & \sum_{x_1\in B_n}\EE  | g_{s,t,w}(\eta,x_1)|^q
+\sum_{r=1}^{n}\sum_{{\bf x}\in G_r(q,n)} \left|\cov \left(
\Pi_{{\bf y(x),w}}, \Pi_{{\bf
z(x),w}}\right)\right|\label{eqn:majaq}\\
&&+\sum_{r=1}^{n}\sum_{{\bf x}\in G_r(q,n)} \left|\EE
\left(\Pi_{{\bf y(x),w}}\right) \EE \left(\Pi_{{\bf
z(x),w}}\right)\right|.\label{eqn:majq}
\end{eqnarray}
Denote $$ V_q(n)=\sum_{x_1\in B_n}\EE  | g_{s,t,w}(\eta,x_1)|^q
+\sum_{r=1}^{n}\sum_{{\bf x}\in G_r(q,n)} \left|\cov \left(
\Pi_{{\bf y(x),w}}, \Pi_{{\bf z(x),w}}\right)\right|.$$  In order
to prove that the expression (\ref{eqn:majq}) is bounded by the
product $\sum_h A_h(n) A_{q-h}(n)$ we make a first summation over
the $\bf x$'s such that ${\bf y(x)}\in B_n^h$. Hence: $$
A_q(n)\leq V_q(n)+\sum_{h=1}^{q-1} A_h(n) A_{q-h}(n). $$ To build
a multi-index ${\bf x}=(x_1,\ldots,x_q)$ belonging to $G_r(q,n)$,
we first fix one of the $|B_n|$ points of $B_n$, say $x_1$.  We
choose a second point $x_2$ with $d(x_1,x_2)= r$. The third point
$x_3$ is in one of the ball with radius $r$ centered in one of the
previous points, and so on\ldots Thus, because the maximal
cardinality of a ball with radius $r$ writes
 $b(r)\le e^{\rho r}$
$$
|G_r(q,n)|\leq |B_n| b(r)2b(r)\cdots (q-1)b(r)
 \leq |B_n|
 (q-1)!2^{q-1}e^{\rho(q-1) r}.$$
 We use lemma
\ref{cor:covx} to deduce:

$$ V_q(n)\leq |B_n| \left(K_\Omega|t-s|+ (q-1)!2^{q-1} \sum_{r=1}^{\infty}
e^{\rho(q-1)r} (c_0qe^{-cr}\wedge K_\Omega |t-s|)\right).
$$
Let $R$ be an integer to be specified, then
$$ V_q(n)\leq |B_n| q!2^{q-1}\left(K_\Omega|t-s|\sum_{r=0}^{R-1} e^{\rho(q-1)r}+
 c_0\sum_{r=R}^{\infty}e^{(\rho (q-1)-c)r}
 \right). $$ Comparing those summations with integrals:
\begin{eqnarray*}
 V_q(n)&\leq & |B_n|q!2^{q-1} \left( \frac{K_\Omega|t-s|}{\rho (q-1)}e^{\rho (q-1)R}
  + \frac{c_0}{c-\rho (q-1)}e^{(\rho (q-1)-c)(R-1)}\right)\\
 & \leq & |B_n| q!2^{q-1}\frac{K_\Omega|t-s|}{c_1}e^{\rho (q-1)R}
 \left( 1  + \frac{c_0}{K_\Omega|t-s|}e^{c-cR}\right),
\end{eqnarray*}
where $c_1=\rho \wedge (c-\rho(2l-1))$. Assume that $(s,t)\in T$
are such that $|t-s|<{c_0e^{c}}/{K_\Omega}$ and choose $R\geq 1$
as the integer such that $ e^{c(R-1)}
\leq\frac{c_0e^{c}}{K_\Omega|t-s|}\leq e^{cR}$.
\begin{equation}\label{eq:deuxsom}
V_q(n)\leq   |B_n|q! \frac{2^qK_\Omega|t-s|e^{2\rho
(q-1)}}{c_1}\left(\frac{c_0}{K_\Omega|t-s|}\right)^{\frac{\rho
(q-1)}c},
\end{equation}
so that $ V_q(n)$ is a function of $q$ that satisfies condition
$({\cal H}_0)$ of Doukhan \& Louhichi \cite{DoukhanLouhichi}. Then
\begin{eqnarray*}
A_{2l}(n) &\leq &\frac{(4l-2)!}{(2l)!(2l-1)!}\left(V_{2l}(n)+V_2(n)^{l}\right)\\
 &\leq &\frac{(4l-2)!(c_0e^{2c})^{\frac{\rho l}c}}{(2l)!(2l-1)!}
\left(\frac{2^{2l}(2l)!(c_0e^{ 2c})^{\frac{\rho (l-1)}c}}{c_1} |B_n|(K_\Omega|t-s|)^{1-\frac{\rho (2l-1)}c}
 \right.\\&&\left.+\left(\frac8 {c_1}\right)^{l}|B_n| ^{l}(K_\Omega|t-s|)^{l-\frac {\rho l} c}
\right),
\end{eqnarray*}
and Proposition \ref{pro:moment} is proved.
\end{proof}
To prove
the tightness of the sequence of processes $N^{B_n}$, we study
its oscillations:
\begin{equation*}
w(\delta,N^{B_n})=\sup_{\|t-s\|_1<\delta}|N_t^{B_n}-N_s^{B_n}|
\end{equation*}
Fix $\varepsilon$ and $\eta$. We have to find $\delta$ and $n_0$
such that for all $n>n_0$ :
\begin{equation*}
\PP (w(\delta,N^{B_n})\geq \varepsilon)\leq \eta
\end{equation*}
Define $n_0$ as the smallest integer such that
$|B_{n_0}|>\delta^{-1-\rho/c}$, then for $n>n_0$, $|t-s|<\delta$,
$l=2$ and $c>3\rho$, Proposition \ref{pro:moment} yields:
\begin{equation*}
\EE (N_t^{B_n}-N_s^{B_n})^{4}\leq C \delta^{2(1-\frac{\rho}{c})}
\end{equation*}
and we now follow the proof in Billingsley \cite{Billingsley} to
conclude.
\subsection{Proof of Theorem \ref{clthitting}}

The proof is close to that of the analogous result in
\cite{ParoissinYcart04}. The convergence in distribution of
$Z_n=(Z_n(t))_{t\ge0}$, where $Z_n(t)=(D_t^{(n)}-|B_n|\cdot
m(t))/\sqrt{ |B_n|}$, does not directly imply the CLT for $T_n$. The
Skorohod-Dudley-Wichura representation theorem is a much stronger
result (see Pollard \cite{Pollard}, section IV.3). It implies that
there exist versions $Z^{*}_n$ of $Z_n$  and non-decreasing
functions $\phi_n$ such that for any fixed $s$ such that for $Z^*$,
limit in distribution of $Z_n$:
\begin{equation*}
\lim_{n \rightarrow \infty} \sup_{0 \leq t \leq s} \left|
Z_n^{*}(t) - Z^*(\phi_n(t)) \right| = 0 \quad a.s.
\end{equation*}
and:
\begin{equation*}
\lim_{n \rightarrow \infty} \sup_{0 \leq t \leq s} \left|
\phi_n(t) - t \right| = 0 \quad a.s.
\end{equation*}
Since $Z^*$ has continuous paths, it is uniformly continuous on
$[0,s]$, and hence:
\begin{equation}\label{unif}
\lim_{n \rightarrow \infty} \sup_{0 \leq t \leq s} \left| Z^*_n(t)
- Z^*(t) \right| = 0 \quad a.s.\;,
\end{equation}
We shall first use (\ref{unif}) to prove that the distributions of
$\sqrt{ |B_n|}(T_n-t_\alpha)$ are a tight sequence. Let $c$ be a
positive constant. On the one hand, if
$D^{(n)}_{t_\alpha+c/\sqrt{ |B_n|}} \geq k(n)$, then $T_n \leq
t_\alpha+c/\sqrt{ |B_n|}$. Thus:
\begin{eqnarray*}
{\PP [ \sqrt{ |B_n|}(T_n - t_\alpha) \leq c]}
 &\geq&{\PP [D^{(n)}_{t_\alpha+c/\sqrt{ |B_n|}}\geq k(n)]} \\[1ex]
 &=&{\PP [Z^*_n(t_\alpha+c/\sqrt{ |B_n|})\geq
 \sqrt{ |B_n|}(\alpha-m(t_\alpha+c/\sqrt{ |B_n|}))+o(1)]} \\[1ex]
 &=&{\PP [Z^*_n(t_\alpha+c/\sqrt{ |B_n|})\geq
 -cm'(t_\alpha)+o(1)]} \\[1ex]
 &=&{\PP [Z^*(t_\alpha)\geq
 -cm'(t_\alpha)]+o(1)}\;,
\end{eqnarray*}
using (\ref{unif}) and the continuity of $Z^*$. Since
$m'(t_\alpha)>0$, we obtain that:
\begin{equation}
\label{liminf} \lim_{c \rightarrow \infty} \liminf_{n \rightarrow
\infty} \PP [
  \sqrt{ |B_n|}(T_n - t_\alpha) \leq c] = 1 .
\end{equation}
On the other hand, we have:
$$
\PP [ \sqrt{ |B_n|}(T_n - t_\alpha) \leq -c]
 =\PP [\exists t \leq t_\alpha-c/\sqrt{ |B_n|}\;,\, Z^*_n(t)\geq
 \sqrt{ |B_n|}(\alpha-m(t))+o(1)] .
$$
But since the function $m$ is increasing, for all $t\leq
t_\alpha-c/\sqrt{ |B_n|}$ we have:
\begin{equation*}
\sqrt{ |B_n|}(\alpha-m(t)) \geq \sqrt{
|B_n|}(\alpha-m(t_\alpha-c/\sqrt{ |B_n|})) = cm'(t_\alpha)+o(1) .
\end{equation*}
Hence:
\begin{eqnarray*}
\displaystyle{\PP [ \sqrt{ |B_n|}(T_n - t_\alpha) \leq -c]}
&\leq&\displaystyle{\PP [\exists t \leq t_\alpha-c/\sqrt{
|B_n|}\,, Z^*_n(t)\geq
 cm'(t_\alpha)+o(1)]} \\[1ex]
&\leq&\displaystyle{ \PP [\exists t \leq t_\alpha\,, Z^*_n(t)\geq
 cm'(t_\alpha)+o(1)]} \\[1ex]
&=&\displaystyle{ \PP [\exists t \leq t_\alpha\,, Z^*(t)\geq
 cm'(t_\alpha)+o(1)]+o(1)} .
\end{eqnarray*}
The process $Z$ being \textsl{a.s.} bounded on any compact set
and $m'(t)$ being positive on $[0,\tau]$, we deduce that:
\begin{equation}
\label{limsup} \lim_{c \rightarrow \infty} \limsup_{n \rightarrow
\infty} \PP [
  \sqrt{ |B_n|}(T_n - t_\alpha) \leq -c] = 0 .
\end{equation}
Now (\ref{liminf}) and (\ref{limsup}) mean that the sequence of
distributions of $(\sqrt{ |B_n|}(T_n-t_\alpha))$ is tight. Hence
to conclude it is enough to check the limit. \vskip 1mm Using
again (\ref{unif}), together with the almost sure continuity of
$Z$ yields:
\begin{eqnarray*}
D^{(n)}_{t_\alpha+c/\sqrt{ |B_n|}}
 & = &  |B_n|m(t_\alpha+u/\sqrt{ |B_n|})
 +\sqrt{ |B_n|}Z^*(t_\alpha+u/\sqrt{ |B_n|})+o(\sqrt{ |B_n|}) \quad a.s. \\[1ex]
 & = &  |B_n|\alpha +u\sqrt{ |B_n|}m'(t_\alpha)
 +\sqrt{ |B_n|}Z^*(t_\alpha)+o(\sqrt{ |B_n|}) \quad a.s.
\end{eqnarray*}
Therefore:
\begin{eqnarray*}
\inf \left\{ u \;; D^{(n)}_{t_\alpha+u/\sqrt{ |B_n|}} \geq k(n) \right\}
  & = & \inf \Big\{ u \;; u\sqrt{ |B_n|}m'(t_\alpha)
 +\sqrt{ |B_n|}Z^*(t_\alpha)+o(\sqrt{ |B_n|}) \geq 0\Big\}\\[1ex]
 & = & - \frac{Z^*(t_\alpha)}{m'(t_\alpha)} +o(1) .
\end{eqnarray*}
The distribution of $-Z^*(t_\alpha)/m'(t_\alpha)$ is normal with
mean $0$ and variance $\sigma^2_\alpha$, hence the result.
\subsection{Proof of Proposition \ref{pro2r}}
Let ${\cal F}_{2,3}$ be
the set of real valued functions $h$
  defined on $\RR$, three times differentiable, such
  that $h(0)=0$, $\|h''\|_{\infty}<+\infty$, and
  $\|h^{(3)}\|_{\infty}<+\infty$. For a function $h\in {\cal
    F}_{2,3}$, we will denote by $b_2$ and $b_3$ the supremum norm of
  its second and third derivatives.
We first need the following lemma.
\begin{lemm}\label{lem1}
Let $h$ be a fixed function of the set ${\cal F}_{2,3}$. Let
$R$ be a fixed and finite subset of $S$. Let $r$ be a fixed
positive real. For any $x\in R$, let $V_x=B(x,r)\cap R$.
Let $(Y_x)_{x\in S}$ be a real valued random field. Suppose that,
for any $x\in S$, $\EE Y_x=0$ and $\EE Y_x^2<+\infty$. Let
$Z(R)=\sum_{x\in R}Y_x$. Then
\begin{eqnarray}\label{cov1}
{\lefteqn{\left|\EE(h(Z(R))) -\Var Z(R)\int_0^1t
\EE(h''\left(tZ(R)\right))dt\right|}} \nonumber\\ &&\leq
\int_0^1\hspace{-1.5mm}\sum_{x\in
R}\left|\Cov\left(Y_x,h'(tZ(V_x^c))\right)\right|dt + 2\sum_{x\in
R}\EE |Y_x||Z(V_x)|\left[b_2\wedge b_3|Z(V_x)|\right] \nonumber\\
&& + b_2 \EE\left|\sum_{x\in
R}\left(Y_xZ(V_x)-\EE(Y_xZ(V_x))\right)\right| + b_2 \sum_{x\in
R}\left|\Cov(Y_x, Z(V^c_x))\right|,
\end{eqnarray}
where $V^c_x=R\setminus V_x$.
\end{lemm}
{\bf{Remark.}} For an independent random field $(Y_x)_{x\in S}$,
fulfilling $\sup_{x\in S}\EE Y_x^4<+\infty$, Lemma \ref{lem1}
applied with $V_x=\{x\}$, ensures $$ \left|\EE(h(Z(R))) -\Var
Z(R)\int_0^1t \EE(h''\left(tZ(R)\right))dt\right|\leq 2 \sum_{x\in
R}\EE |Y_x|^2\left(b_2\wedge b_3|Y_x|\right) +
b_2\sqrt{|R|}\sup_{x\in S}\|Y_x^2\|_2 . $$
{\bf{Proof of Lemma \ref{lem1}.}} We have,
\begin{eqnarray*}\label{d1}
{\lefteqn{h(Z(R)) = Z(R)\int_0^1 h'(tZ(R))dt  =
\int_0^1\left(\sum_{x\in R}Y_xh'(tZ(R))\right)dt}} {\nonumber}\\
&& = \int_0^1\left(\sum_{x\in R}Y_xh'(tZ(V_x^c))\right)dt +
\int_0^1\left(\sum_{x\in
R}Y_x\left(h'(tZ(R))-h'(tZ(V_x^c))-tZ(V_x)h''(tZ(R))\right)\right)dt{\nonumber}\\
&& + \sum_{x\in R}Y_xZ(V_x)\int_0^1th''(tZ(R))dt - \sum_{x\in
R}\EE\left(Y_xZ(V_x)\right)\int_0^1th''(tZ(R))dt\\ && + \sum_{x\in
R}\EE\left(Y_xZ(V_x)\right)\int_0^1th''(tZ(R))dt
-\sum_{x\in R}\EE\left(Y_xZ(R)\right)\int_0^1th''(tZ(R))dt {\nonumber}\\
&& + \sum_{x\in
R}\EE\left(Y_xZ(R)\right)\int_0^1th''(tZ(R))dt{\nonumber}.
\end{eqnarray*}
We take expectation in the last equality. The obtained formula,
together with the following estimations, proves Lemma \ref{lem1}.
\begin{eqnarray*}
{\lefteqn{\left|h'(tZ(R))-h'(tZ(V_x^c))-tZ(V_x)h''(tZ(R))\right|}}
\\ && \leq
\left|h'(tZ(R))-h'(tZ(V_x^c))-tZ(V_x)h''(tZ(V_x^c))\right|+
|Z(V_x)||h''(tZ(R))-h''(tZ(V_x^c))| \\ && \leq 2
|Z(V_x)|\left(b_2\wedge b_3|Z(V_x)|\right).\qquad\Box
\end{eqnarray*}

\vskip 3mm\noindent
Our purpose now is to control the right hand side of the bound
(\ref{cov1}) for a random field $(Y_x)_{x\in S}$ fulfilling the
covariance inequality (\ref{slcov}) and the requirements of
Proposition \ref{pro2r}.
\begin{coro}\label{cor1}
Let $h$ be a fixed function of the set ${\cal F}_{2,3}$. Let
$R$ be a finite subset of $S$.  For any $x\in R$ and for any
positive real $r$, let $V_x=B(x,r)\cap R$. Let $(Y_x)_{x\in
S}$ be a real valued random field, fulfilling the covariance
inequality (\ref{slcov}). Suppose that, for any $x\in S$, $\EE
Y_x=0$ and $\sup_{x\in S}\|Y_x\|_{\infty}< M$, for some positive
real $M$. Recall that $Z(R)=\sum_{x\in R}Y_x$. Then, for
any $\delta>0$, there exists a positive constant $C({\delta},M)$
independent of $R$, such that
\begin{eqnarray*}
{\lefteqn{\sup_{h\in {\cal F}_{2,3}}\left|\EE(h(Z(R))) -\Var
Z(R)\int_0^1t \EE(h''\left(tZ(R)\right))dt\right|}} \\ && \leq
C({\delta},M)\left\{ b_2 |R|e^{-\delta r}+ b_3 |R| \kappa_r
+ b_2|R|^{1/2}\kappa_r\left(\sum_{k=[3r]}^{\infty}\kappa_k e^{-\delta(k-2r)}\right)^{1/2}\right.\\
&& \left. +
b_2|R|^{1/2}\kappa_{3r}\left(\sum_{k=1}^{[3r]+1}e^{-\delta
k}\kappa_k\right)^{1/2}\right\},
\end{eqnarray*}
recall that $\sup_{x\in S}|B(x,n)|\leq \kappa_n$.
\end{coro}
{\bf{Proof of Corollary \ref{cor1}}}
\\
We have $$V_x^c=\{y\in S,\ \ d(x,y)\geq r\}\cap R. $$ Hence
$$d(\{x\}, V^c_x)\geq r. $$ The last bound together with
(\ref{slcov}), proves that
\begin{eqnarray}\label{t1}
\sum_{x\in R}\left|\Cov\left(Y_x,h'(tZ(V_x^c))\right)\right|&\leq
& C_{\delta} b_2 \sum_{x\in R}(|V_x^c|\wedge 1)e^{-\delta d(\{x\},
V^c_x)} \nonumber
\\ & \leq &  C_{\delta} b_2 |R|e^{-\delta r}.
\end{eqnarray}
In the same way, we prove that
\begin{eqnarray}\label{t2}
b_2 \sum_{x\in R}\left|\Cov(Y_x, Z(V^c_x))\right| & \leq &
C_{\delta} b_2 |R|e^{-\delta r}.
\end{eqnarray}
Now
\begin{eqnarray}\label{t3}
\sum_{x\in R}\EE |Y_x||Z(V_x)|\left(b_2\wedge b_3|Z(V_x)|\right)
&\leq & b_3 M |R| \sup_{x\in S} \EE|Z(V_x)|^2 {\nonumber}\\
&\leq &
 b_3 M |R|\kappa_r \sup_{y\in S}
\sum_{z\in S}|\Cov(Y_y,Y_z)|
\end{eqnarray}
The last bound is obtained since $|V_x|\leq \kappa_r$ and
$\sup_{y\in S} \sum_{z\in S}|\Cov(Y_y,Y_z)|<\infty$ (the proof of
the last inequality is done along the same lines as that of
Proposition \ref{slpro2}) .
\\
It remains to control $$\EE\left|\sum_{x\in
R}\left(Y_xZ(V_x)-\EE(Y_xZ(V_x))\right)\right|.$$ For this, we
argue as Bolthausen \cite{Bolthausen82}. We have
\begin{eqnarray*}
\EE\left|\sum_{x\in
R}\left(Y_xZ(V_x)-\EE(Y_xZ(V_x))\right)\right|^2 &=&
\Var(\sum_{x\in R}Y_xZ(V_x)) \\ & =& \sum_{x\in R}\sum_{y\in
R}\Cov(Y_xZ(V_x), Y_yZ(V_y)).
\end{eqnarray*}
Hence, since $V_x\subset B(x,r)$,
\begin{eqnarray}\label{i1}
\EE\left|\sum_{x\in
R}\left(Y_xZ(V_x)-\EE(Y_xZ(V_x))\right)\right|^2\leq \sum_{x\in
R}\sum_{x'\in B(x,r)}\sum_{y\in R}\sum_{y'\in
B(y,r)}\left|\Cov(Y_xY_{x'}, Y_yY_{y'})\right|.
\end{eqnarray}
We have,
\begin{eqnarray}\label{second}
 \left|\Cov(Y_xY_{x'}, Y_yY_{y'})\right|
 \leq  \left|\Cov(Y_xY_{x'}, Y_yY_{y'})\right|\II_{d(x,y)\geq 3r}+
\left|\Cov(Y_xY_{x'}, Y_yY_{y'})\right|\II_{d(x,y)\leq 3r}.
\end{eqnarray}
We begin by controlling the first term. The covariance inequality
(\ref{slcov}) together with some elementary estimations, ensures
\begin{eqnarray*}
\left|\Cov(Y_xY_{x'}, Y_yY_{y'})\right|\II_{d(x,y)\geq 3r} & \leq
& \sum_{k=[3r]}^{\infty}\left|\Cov(Y_xY_{x'},
Y_yY_{y'})\right|\II_{k\leq d(x,y)< k+1} \\ & \leq & 2 M^2
C_{\delta}\sum_{k=[3r]}^{\infty}e^{-\delta
d(\{x,x'\},\{y,y'\})}\II_{k\leq d(x,y)< k+1}\\ & \leq & 2 M^2
C_{\delta}\sum_{k=[3r]}^{\infty}e^{-\delta(k-2r)}\II_{ d(x,y)<
k+1},
\end{eqnarray*}
the last bound is obtained since, for any $x'\in B(x,r)$ and
$y'\in B(y,r)$, we have, $$ d(\{x,x'\},\{y,y'\})+2r \geq
d(\{x,x'\},\{y,y'\})+ d(x,x')+ d(y,y')\geq d(x,y).$$ Hence,
\begin{eqnarray}\label{i2}
{\lefteqn{\sum_{x\in R}\sum_{x'\in B(x,r)}\sum_{y\in R}\sum_{y'\in
B(y,r)}\left|\Cov(Y_xY_{x'}, Y_yY_{y'})\right|\II_{d(x,y)\geq
3r}}}\nonumber \\ &&\leq 2 M^2
C_{\delta}\kappa_r^2\sum_{k=[3r]}^{\infty}\sum_{x\in R}\sum_{y\in
R}e^{-\delta(k-2r)} \II_{y\in B(x,k+1)}\nonumber
\\
&&\leq 2 M^2
C_{\delta}|R|\kappa_r^2\sum_{k=[3r]}^{\infty}\kappa_{k+1}e^{-\delta(k-2r)}.
\end{eqnarray}
We now control the second term in (\ref{second}). Inequality
(\ref{slcov}) and the fact that\\ $d(\{x\},\{x',y,y'\})\leq
d(\{x\},\{x'\})$, ensure
\begin{eqnarray*}
{\lefteqn{\left|\Cov(Y_xY_{x'},
Y_yY_{y'})\right|\II_{d(x,y)\leq 3r}}}\\
 &&\leq \left|\Cov(Y_x,
Y_{x'}Y_yY_{y'})\right|\II_{d(x,y)\leq 3r}+
\left|\Cov(Y_x,Y_{x'})\right|\left|\Cov(Y_y,Y_{y'})\right|\II_{d(x,y)\leq
3r}\\ && \leq 2M^2C_{\delta}e^{-\delta
d(\{x\},\{x',y,y'\})}\II_{d(x,y)\leq 3r}.
\end{eqnarray*}
We deduce, using the last bound, that
\begin{eqnarray}\label{bff}
{\lefteqn{\left|\Cov(Y_xY_{x'},
Y_yY_{y'})\right|\II_{d(x,y)\leq 3r}}}\nonumber\\
 &&\leq \sum_{k=1}^{[3r]+1}\left|\Cov(Y_xY_{x'},
Y_yY_{y'})\right|\II_{d(x,y)\leq 3r}\II_{k-1\leq
d(\{x\},\{x',y,y'\})< k}\nonumber\\ && \leq
2M^2C_{\delta}\sum_{k=1}^{[3r]+1}e^{-\delta (k-1)}\II_{d(x,y)\leq
3r}\II_{d(\{x\},\{x',y,y'\})< k}.
\end{eqnarray}
We have $$\II_{d(\{x\},\{x',y,y'\})\leq k}\leq
\II_{d(\{x\},\{x'\})\leq k}+ \II_{d(\{x\},\{y\})\leq
k}+\II_{d(\{x\},\{y'\})\leq k}.$$ Hence, we check that,
\begin{eqnarray}\label{bf}
\sum_{x\in R}\sum_{x'\in B(x,r)}\sum_{y\in R}\sum_{y'\in
B(y,r)}\II_{d(x,y)\leq 3r}\II_{d(\{x\},\{x',y,y'\})\leq k} \leq
3|R|\kappa_{3r}^2\kappa_k.
\end{eqnarray}
We obtain combining (\ref{bff}) and (\ref{bf}),
\begin{multline}
 \sum_{x\in R}\sum_{x'\in B(x,r)}\sum_{y\in R}\sum_{y'\in
B(y,r)}\left|\Cov(Y_xY_{x'}, Y_yY_{y'})\right|\II_{d(x,y)\leq
3r}\\
\label{i3} \leq
6e^{\delta}M^2C_{\delta}|R|\kappa_{3r}^2\sum_{k=1}^{[3r]+1}e^{-\delta
k}\kappa_k.
\end{multline}
We collect the bounds (\ref{i1}), (\ref{i2}) and (\ref{i3}), we
obtain,
\begin{eqnarray}\label{t4}
{\lefteqn{\EE\left|\sum_{x\in
R}\left(Y_xZ(V_x)-\EE(Y_xZ(V_x))\right)\right|}}\nonumber\\
&&\leq
C(\delta,M)|R|^{1/2}\left\{\kappa_r\left(\sum_{k=[3r]}^{\infty}\kappa_{k+1}e^{-\delta(k-2r)}\right)^{1/2}
+ \kappa_{3r} \left(\sum_{k=1}^{[3r]+1}e^{-\delta
k}\kappa_k\right)^{1/2}\right\}.
\end{eqnarray}
Finally, the bounds (\ref{t1}), (\ref{t2}), (\ref{t3}),
(\ref{t4}), together with Lemma \ref{lem1} prove Corollary
\ref{cor1}. \ \ \ \  $\Box$
\vskip 3mm\noindent
 {\bf{End of the proof of Proposition \ref{pro2r}.}}
We apply Corollary \ref{cor1} to the real and imaginary parts of
the function $x\rightarrow \exp(iux/{\sqrt{|B_n|}})-1$. Those
functions belong to the set ${\cal F}_{2,3}$, with
$b_2=\frac{\textstyle u^2}{\textstyle |B_n|}$ and
$b_3=\frac{\textstyle |u|^3}{\textstyle |B_n|^{3/2}}$.
\\
We obtain, noting by $\phi_n$ the characteristic function of the
normalized sum $Z(B_n)/{\sqrt{|B_n|}}$,
\begin{eqnarray*}
{\lefteqn{\left|\phi_n(u)-1+\frac{\Var Z(B_n)}{|B_n|}u^2
\int_0^1t\phi_n(tu)dt\right|}}
\\ && \leq C({\delta},M,u)\left\{
e^{-\delta r}+ \frac{\kappa_r}{\sqrt{|B_n|}} +
\frac{\kappa_r}{\sqrt{|B_n|}}\left(\sum_{k=[3r]}^{\infty}\kappa_k e^{-\delta(k-2r)}\right)^{1/2}\right.\\
&& \left.
+\frac{\kappa_{3r}}{\sqrt{|B_n|}}\left(\sum_{k=1}^{[3r]+1}e^{-\delta
k}\kappa_k\right)^{1/2}\right\}.
\end{eqnarray*}
Let $\delta$ be a fixed positive real such that
$\delta>12\rho$, recall that
$$\sup_{x\in S}|B(x,r)|\leq 2 e^{r\rho}=:\kappa_r.$$ Hence
\begin{eqnarray*}
{\lefteqn{\left|\phi_n(u)-1+\frac{\Var Z(B_n)}{|B_n|}u^2
\int_0^1t\phi_n(tu)dt\right|}}
\\ && \leq C(\delta,M,u)\left\{
e^{-\delta r}+
\frac{e^{r\rho}}{\sqrt{|B_n|}}+\frac{e^{(\rho+\delta)r}}{\sqrt{|B_n|}}\left(\sum_{k=[3r]}^{\infty}e^{-(\delta-\rho)k}\right)^{1/2}
 +\frac{e^{3\rho
r}}{\sqrt{|B_n|}}\left(\sum_{k=1}^{[3r]+1}e^{-(\delta-\rho)
k}\right)^{1/2}\right\}\\ &&\leq
C(M,\rho,\delta,u)\left(e^{-\delta
r}+\frac{e^{3r\rho}}{\sqrt{|B_n|}}+
\frac{e^{-(\delta-5\rho)r/2}}{\sqrt{|B_n|}}\right).
\end{eqnarray*}
For a suitable choice of the sequence $r$ (for example we can take
$r=\frac{2}{\delta}\ln|B_n|$), the right hand side of the last
bound tends to $0$ an $n$ tends to infinity:
\begin{equation}\label{limf}
\lim_{n\rightarrow \infty}\left|\phi_n(u)-1+\frac{\Var
Z(B_n)}{|B_n|}u^2 \int_0^1t\phi_n(tu)dt\right|=0.
\end{equation}
We now need the following lemma.
\begin{lemm}\label{lem2} Let $\sigma^2$ be a positive real. Let $(X_n)$
be a sequence of real valued random variables such that
$\sup_{n\in \NN}\EE X_n^2<+\infty$. Let $\phi_n$ be the
characteristic function of $X_n$. Suppose that for any $u\in \RR$,
\begin{equation}\label{cara}
\lim_{n\rightarrow
+\infty}\left|\phi_n(u)-1+\sigma^2\int_0^ut\phi_n(t)dt\right|=0.
\end{equation}
Then, for any $u\in \RR$, $$\lim_{n\rightarrow
+\infty}\phi_n(u)=\exp(-\frac{u^2\sigma^2}{2}). $$
\end{lemm}
{\bf{Proof of Lemma \ref{lem2}.}}
 Lemma \ref{lem2} is  a variant of Lemma 2 in
Bolthausen \cite{Bolthausen82}. The  Markov inequality and the
condition $\sup_{n\in \NN}\EE X_n^2<+\infty$ imply that the
sequence $(\mu_n)_{n\in \NN}$ of the laws of $(X_n)$ is tight.
Theorem 25.10 in Billingsley \cite{Billingsley} proves the
existence of a subsequence $\mu_{n_k}$ and a probability measure
$\mu$ such that $\mu_{n_k}$ converges weakly to $\mu$ as $k$ tends
to infinity. Let $\phi$ be the characteristic function of $\mu$.
We deduce from (\ref{cara}) that, for any $u\in \RR$, $$
\phi(u)-1+\sigma^2\int_0^ut\phi(t)dt=0, $$ or equivalently, for
any $u\in \RR$, $$ \phi'(u)+\sigma^2u\phi(u)=0. $$ We obtain,
integrating the last equation, that for any $u\in \RR$, $$
\phi(u)= \exp(-\frac{\sigma^2 u^2}{2}). $$ The proof of Lemma
\ref{lem2} is completed by using Theorem 25.10 in Billingsley
\cite{Billingsley} and its corollary. \,\,\, $\Box$ \vskip 1mm
Proposition \ref{pro2r} follows from (\ref{limff}), (\ref{limf})
and Lemma \ref{lem2}.\ \ \ $\Box$ \vskip 3mm
\subsection{Proof of Proposition \ref{slpro2}.}
We deduce from
(\ref{slcov}) that for any positive real $\delta$ there
exists a positive constant $C_{\delta}$ such that for different
sites $x$ and $y$ of $S$,
\begin{equation}\label{slc1}
\left|\Cov(Y_x,Y_y)\right|\leq C_{\delta} e^{-\delta d(x,y)}.
\end{equation}
Hence, the first conclusion of Proposition \ref{slpro2} follows
from the bound (\ref{slc1}), together with the following
elementary calculations, for $\rho<\delta$,
\begin{eqnarray}\label{slcf}
\sum_{z\in S}|\Cov(Y_0,Y_{z})| &\leq & C_{\delta}  \sum_{z\in
S}\exp(-\delta d(0,z)){\nonumber}\\
& \leq & C_{\delta}\sum_{z\in S}\sum_{r=0}^{\infty}\exp(-\delta
d(0,z))\II_{r \leq d(0,z)< r+1} {\nonumber}\\ & \leq &
C_{\delta}\sum_{r= 0}^{\infty}\exp(-\delta
r)\sum_{z\in S}\II_{d(0,z)< r+1}{\nonumber}\\
& \leq & C_{\delta}\sum_{r=0}^{\infty}\exp(-\delta
r)|B(0,r+1)|{\nonumber}\\ & \leq & C(\delta,\rho)\sum_{r=
0}^{\infty}\exp(-(\delta -\rho)r),
\end{eqnarray}
where $C(\delta,\rho)$ is a positive constant depending  on
$\delta$ and $\rho$.
\\
We now prove the second part of Proposition \ref{slpro2}. Thanks
to (\ref{slc}), we can find a sequence $u=(u_n)$ of positive real
numbers such that
\begin{equation}\label{sllimd}
\lim_{n\rightarrow +\infty}u_n=+\infty, \, \lim_{n\rightarrow
+\infty} \frac{|\partial B_n|}{|B_n|}\exp(\rho u_n)=0.
\end{equation}
Let $(\partial _{u}B_n)_n$ be the sequence of subsets of $S$
defined by
$$\partial _{u}B_n=\{s\in B_n\,:\, d(s,\partial B_n)< u_n\}. $$
The bound (\ref{borneboule}) gives
$$ |\partial _{u}B_n|\leq  2 |\partial B_n| e^{u_n\rho},
$$
which together with the suitable choice of the sequence  $(u_n)$
ensures
\begin{equation}\label{sllimt3}
\lim_{n\rightarrow +\infty}\frac{|\partial _{u}B_n|}{|B_n|}=0,
\end{equation}
we shall use this fact below without further comments. Let
$B^{u}_n=B_n\setminus \partial _{u}B_n$. We decompose the quantity
$\Var\,S_n$ as in Newman \cite{Newman80}:
\begin{eqnarray*}\label{sldeco}
\frac{1}{|B_n|}\Var\, S_n & = & \frac{1}{|B_n|}\sum_{x\in
B_n}\sum_{y\in B_n}\Cov\left(Y_x,Y_y\right) = T_{1,n} + T_{2,n}
+T_{3,n},
\end{eqnarray*}
where
\begin{eqnarray*}
T_{1,n} & = & \frac{1}{|B_n|}\sum_{x\in B^{u}_n}\,\,\sum_{y\in
B_n\setminus B(x,u_n)}\Cov\left(Y_x,Y_y\right),\\
T_{2,n} & = & \frac{1}{|B_n|}\sum_{x\in B^{u}_n}\,\,\sum_{y\in
B_n\cap B(x,u_n)}\Cov\left(Y_x,Y_y\right),\\ T_{3,n} & = &
\frac{1}{|B_n|}\sum_{x\in \partial_u B_n}\sum_{y\in
B_n}\Cov\left(Y_x,Y_y\right).
\end{eqnarray*}
{\bf{Control of $T_{1,n}$.}} We have, since $|B^{u}_n| \leq |B_n|$
and applying (\ref{slc1})
\begin{eqnarray}\label{slt1bisbis}
|T_{1,n}| &\leq & \sup_{x\in S} \sum_{y\in S\setminus B(x,u_n)
}\left|\Cov(Y_x,Y_y)\right| \leq  C_\delta \sup_{x\in S}
\sum_{y\in S\setminus B(x,n)}\exp(-\delta d(x,y)).
\end{eqnarray}
For any fixed $x\in S$, we argue as for (\ref{slcf}) and we obtain
for $\rho<\delta$,
\begin{equation}\label{slt1bis}
\sum_{y\in S\setminus B(x,n)}\exp(-\delta d(x,y)) \leq
C(\delta)\sum_{r= [u_n]}^{\infty}\exp(-(\delta-\rho) r) \leq
C(\delta,\rho)\exp(-(\delta-\rho)u_n)
\end{equation}
 We obtain, collecting (\ref{slt1bisbis}), (\ref{slt1bis}) together with
the first limit in (\ref{sllimd}) :
\begin{equation}\label{sllimT1}
\lim_{n\rightarrow +\infty} T_{1,n}=0.
\end{equation}
\\
{\bf{Control of $T_{3,n}$.}} We obtain using (\ref{slc1}) :
\begin{eqnarray}\label{slt3bis}
|T_{3,n}| &\leq & \frac{|\partial_u B_n|}{|B_n|} \sup_{x\in S}
\sum_{y\in S}\left|\Cov(Y_x,Y_y)\right|.
\end{eqnarray}
The last bound, together with the limit (\ref{sllimt3}) 
gives
\begin{equation}\label{sllimT3}
\lim_{n\rightarrow +\infty}T_{3,n}=0.
\end{equation}
 \\ {\bf{Control of $T_{2,n}$.}}
 We deduce using the following implication, if $x\in
B^{u}_n$ and $y$ is not belonging to $B_n$ then $d(x,y)\geq u_n$,
that
\begin{eqnarray*}
T_{2,n} & = & \frac{1}{|B_n|}\sum_{x\in B^{u}_n}\,\,\sum_{y\in
B(x,u_n)}\Cov(Y_x,Y_y)
\end{eqnarray*}
We claim that,
\begin{equation}\label{ccc}
\sum_{y\in B(x,u_n)}\Cov(Y_x,Y_y) = \sum_{z\in
B(0,u_n)}\Cov\left(Y_0,Y_z\right),
\end{equation}
in fact, since the graph ${\cal G}$ is transitive, there exits an
automorphism $a_x$, such that $a_x(x)=0$ ($0$ is a fixed vertex in
$S$). Equality (\ref{slstationarity}) gives
$$
\sum_{y\in B(x,u_n)}\Cov(Y_x,Y_y)=\sum_{y\in
B(x,u_n)}\Cov(Y_0,Y_{a_x(y)}).
$$
Now, Lemma 1.3.2 in Godsil and Royle \cite{GodsilRoyle} yields that $d(x,y)=
d(a_x(x), a_x(y))=d(0,a_x(y))$. From this we deduce that $y\in
B(x,u_n)$ if and only if $a_x(y)\in B(0,u_n)$. From above, we
conclude that,
$$
\sum_{y\in B(x,u_n)}\Cov(Y_x,Y_y)=\sum_{a_x(y)\in
B(0,u_n)}\Cov(Y_0,Y_{a_x(y)})=\sum_{z\in B(0,u_n)}\Cov(Y_0,Y_{z}),
$$
which proves (\ref{ccc}). Consequently,
\begin{eqnarray*}
T_{2,n} & = &  \frac{|B^{u}_n|}{|B_n|}\sum_{z\in B(0,u_n)
}\Cov(Y_0,Y_z).
\end{eqnarray*}
The last equality together with the first limit in (\ref{sllimd})
and (\ref{sllimt3}), ensures
\begin{equation}\label{sllimT2}
\lim_{n\rightarrow +\infty} T_{2,n}= \sum_{z\in S}\Cov(Y_0,Y_z).
\end{equation}
The second conclusion of Proposition \ref{slpro2} is proved by
collecting the limits (\ref{sllimT1}), (\ref{sllimT3}) and
(\ref{sllimT2}). $\Box$
\paragraph{Acknowledgements.} We wish to thank Professor Mathew Penrose for
his important remarks which helped us to derive the present version
of this work. He mentioned an error in a previous draft for this
work, see the Remark following Theorem \ref{cltips2}. We also thank
David Coupier for his precious comments.
\bibliographystyle{plain}

\bibliography{clti.bbl}

\end{document}